\documentclass[british,a4paper,10pt]{article}
\usepackage{amssymb}
\usepackage{amsmath}
\usepackage{subfigure}
\usepackage{slashed}
\usepackage{amsmath}
\usepackage[normalem]{ulem}
\usepackage{graphicx,psfrag}
\usepackage{mathrsfs}
\usepackage{amsfonts}
\usepackage{multirow}
\usepackage{hyperref}
\usepackage{placeins}
\usepackage{amsthm}
\usepackage{latexsym}
\usepackage[dvips]{epsfig}
\usepackage{enumitem}  
\usepackage{soul}
\usepackage{ulem}
\usepackage{graphicx}
\usepackage{amsthm}
\usepackage{latexsym}
\usepackage[dvips]{epsfig}
\usepackage[utf8]{inputenc}
\usepackage{babel}
\usepackage{multicol}
\usepackage{caption}
\usepackage{color}
\usepackage{comment}
\usepackage[dvipsnames]{xcolor}
\usepackage{wasysym}
\usepackage{mathrsfs}
\usepackage{orcidlink}
\usepackage{bm}
\usepackage{authblk}
\usepackage{slashed}
\usepackage{yhmath} 
\usepackage{stmaryrd}

\usepackage{pifont}
\usepackage{mdframed}
\newcommand{\stkout}[1]{\ifmmode\text{\sout{\ensuremath{#1}}}\else\sout{#1}\fi}

\theoremstyle{plain}
\newtheorem{proposition}{Proposition}

\newtheorem{assumption}{Assumption}

\newtheorem{remark}{Remark}
\def\bma{{\bm a}}

\def\bmn{{\bm n}}
\def\bmeta{{\bm \eta}}
\def\bmg{{\bm g}}

\def\bmf{{\bm f}}

\def\bmK{{\bm K}}

\def\dotnabla{{\mathring{\nabla}}}
\def\dotg{{\mathring{g}}}

\def\dotXi{{\mathring{\Xi}}}

\def\dotPhi{{\mathring{\Phi}}}

\usepackage{graphicx}
\newcommand\smallO{
  \mathchoice
    {{\scriptstyle\mathcal{O}}}
    {{\scriptstyle\mathcal{O}}}
    {{\scriptscriptstyle\mathcal{O}}}
    {\scalebox{.7}{$\scriptscriptstyle\mathcal{O}$}}
  }

\usepackage[symbol]{footmisc}

\newcounter{mnotecount}

\newcommand{\mnotex}[1]
{\protect{\stepcounter{mnotecount}}$^{\mbox{\footnotesize $\bullet$\themnotecount}}$ 
\marginpar{
\raggedright\tiny\em
$\!\!\!\!\!\!\,\bullet$\themnotecount: #1} }

\allowdisplaybreaks
\begin{document}

\title{\textbf{Numerical evolutions of the linearised conformal Einstein field equations
  in the inversion-Minkowski spacetime}}
\author[1]{Christian Peterson \orcidlink{0000-0003-4842-1368} \footnote{E-mail address:{\tt christian.peterson@tecnico.ulisboa.pt}}}
\author[1]{Edgar Gasper\'in \orcidlink{0000-0003-1170-5121}  \footnote{E-mail address:{\tt edgar.gasperin@tecnico.ulisboa.pt}}}
\author[1]{Alex Va\~n\'o-Vi\~nuales \orcidlink{0000-0002-8589-006X} \footnote{E-mail address:{\tt alex.vano.vinuales@tecnico.ulisboa.pt}}}
  \affil[1] { CENTRA, Departamento de F\'isica, Instituto Superior
    T\'ecnico IST, Universidade de Lisboa UL, Avenida Rovisco Pais 1,
    1049 Lisboa, Portugal }

  \maketitle

  \vspace{-4mm}
  
\begin{abstract}
        Numerical evolutions of a system of equations close to null
        infinity in the geometric background of the 
        \emph{inversion-Minkowski spacetime} are performed. The evolved
        equations correspond to the linearisation of a second order
        metric formulation of the conformal Einstein field equations
        (CEFEs). These linear equations were first presented in
        \cite{FenGas23} and the main purpose of this paper is to
        illustrate, through numerical evolutions, the scri-fixing
        technique via gauge source functions introduced
        \cite{FenGas23} for the linearised CEFEs. A
        comparison of evolutions using scri-fixing gauge sources and
        trivial gauge source function in spherical and axial symmetry is
        presented.
\end{abstract}


\section{Introduction}

The original notion of gravitational waves can be traced back to the
linearisation of the Einstein field equations and the well-known
quadrupole formula ---for a historical overview see \cite{HilNur16,
  BieGarYun17}. However, it took several decades for a complete
understanding gravitational radiation at the non-linear level to be
achieved \cite{Bon62, Sac62, NewPen62}.  The reason is that the
non-linear notion of gravitational radiation requires to be able to
evaluate certain quantities at null-infinity $\mathscr{I}$
\cite{Ger77, Ash81, Fer24, Ash14} ---the mathematical idealisation of
the asymptotic region of spacetime. In turn, $\mathscr{I}$ is formally
defined not at the level of the physical spacetime
$(\tilde{\mathcal{M}},\tilde{\bmg})$ but rather through a conformally
related manifold with metric $(\mathcal{M}, \bmg)$ where
$\bmg=\Xi^2\tilde{\bmg}$ which we call \emph{unphysical spacetime}
---see \cite{Pen64} for the origin of this terminology. Then,
$\mathscr{I}$ is identified with the points in $\mathcal{M}$ for which
$\Xi=0$ but $\mbox{d}\Xi \neq 0$.  Given that the Einstein field
equations are not conformally invariant obtaining an initial value
formulation that includes $\mathscr{I}$ is not trivial.  Yet, nowadays
there are different proposals to include $\mathscr{I}$ in the initial
value formulation of general relativity. One can classify some of
these proposals by analysing how the conformal factor $\Xi$ is
treated. The first approach is to take $\Xi$ ---or some other
boundary-defining function--- as a non-dynamical quantity and evolve
the unphysical metric $\bmg$ ---or other equivalent set of rescaled
variables--- using hyperboloidal foliations.  Some examples of
formulations of the Einstein field equations where this approach is
followed are \cite{Zen08, Van15, Van23, HilHarBug16, PetGauVanHil24}
and, although they are different in several important details they
share essentially the core idea described above and we will refer to
them as the \emph{formally singular hyperboloidal approach}.  Its
most evident advantage is that it allows to set up the Einstein
equations for hyperboloidal evolution in a way similar to current
Numerical Relativity simulations of merging binaries targeting
astrophysical applications.  Another positive point is that the
prescribed conformal factor $\Xi$ allows to set the location of
$\mathscr{I}$ at a fixed coordinate location (usually the outer
boundary of the domain).  The second approach to the inclusion of
$\mathscr{I}$ is to consider $\Xi$ as a dynamical quantity, with its
evolution governed directly by the Einstein field equations.
Promoting $\Xi$ to the level of an evolved field forces one to
introduce the curvature as a further unknown which then results in a
set of equations, which although equivalent to the Einstein field
equations when $\Xi=1$, look very different to standard formulations
of general relativity.  Following this general strategy one arrives at
the \emph{conformal Einstein field equations} (CEFEs) originally
introduced in \cite{Fri81} by H. Friedrich ---see \cite{Val16} for a
comprehensive discussion. The basic advantage of the CEFEs is that, in
contrast with the formally singular hyperboloidal approach, they are
\emph{formally regular} in the sense that $\Xi^{-1}$ terms do not
appear in the equations.  Despite the fact that there are already a
number of numerical and mathematical results obtained with the CEFEs
\cite{Fri81, Fri02, GasVal17, MinVal23, DouFra16, Hub99,
  Hub01,FraSte21, FraGooSte23} in general, they remain to some extent
relatively unexploited for physical applications as compared with
other more standard formulations.  Given that the linearisation of the
Einstein field equations has been the starting point of several
important areas aimed to astrophysical applications such as as the
self-force program, quasinormal modes and black hole spectroscopy, it
is natural to pursue analogous calculations using the CEFEs to
evaluate quantities of physical interest at $\mathscr{I}$.
Furthermore, the relevance of the inclusion of $\mathscr{I}$ in these
calculations is evidenced by the success of the hyperboloidal approach
for linear perturbation theory \cite{AnsMac16, ZenNunHus09,
  JarMacShe21, MacZen24, MacBouPouUpt24}.  A fundamental difference
between the hyperboloidal approach in the study of linear equations
and that of the non-linear equations is that in the former case the
equations are regular while in the latter case, formally singular.
Nevertheless, numerical evidence suggest that these formally singular
terms (composed of divergent factors multiplied by fields that decay
implicitly) in fact acquire finite limits. Hence, with these
motivations, in \cite{FenGas23} the CEFEs were linearised using a
particular form that is closer to traditional formulations of the
Einstein field equations used in Numerical Relativity: wave equations
with gauge source functions (generalised harmonic gauge).  Therefore,
\cite{FenGas23} represents the linear version of the non-linear
formulation \cite{CarHurVal, Pae13} which seems to be the hyperbolic
reduction of the CEFEs which more closely resembles standard
formulations used in Numerical Relativity.  One of the lessons learned
in \cite{FenGas23} is that, at least at the linear level, the freedom
in the gauge source functions can be used to fix the conformal factor
$\Xi$ in an, essentially, ad-hoc way.  A similar scri-fixing through
gauge source functions in the non-linear case would open the
possibility of using the coordinates and gauges well understood in the
formally singular hyperboloidal approach \cite{Van15, Van23, VanVal24,
  MacZen24} but with the formally regular equations that the CEFEs
provide. Whether this is in fact possible, and its relation to
Zenginoğlu's scri-fixing \cite{Zen07a, Zen08} and Frauendiener's
scri-freezing shift gauges \cite{Fra98a} will be explored elsewhere.
To test these ideas numerically in a relatively simple arena, in this
paper we perform numerical evolutions of the linearised metric and
second order CEFEs derived in \cite{FenGas23}.

\section{Geometric set-up and formulation}

Although the purpose of this article is to perform numerical
evolutions of the linearised CEFEs presented in \cite{FenGas23}, to
have an abridged but self-contained discussion, in this section we
briefly revisit the non-linear CEFEs and its linearisation.

\subsection{The conformal Einstein field equations}
\label{Sec:CFEs}

\medskip

The conformal Einstein field equations (CEFEs) were originally
introduced in 1981 in \cite{Fri81}. This reformulation of the Einstein
field equations is based on Penrose's idea of conformal
compactification in which $(\tilde{\mathcal{M}}, \tilde{\bmg})$ is the
physical spacetime that satisfies the Einstein field equations and one
considers a conformally related manifold with metric
$(\mathcal{M},\bmg, \Xi)$ where $\bmg=\Xi^2\tilde{\bmg}$ that is
called the unphysical spacetime.  If the physical spacetime satisfies
the vacuum Einstein field equations $\tilde{R}_{ab}=\lambda
\tilde{g}_{ab}$ then the CEFEs read

\begin{subequations}\label{CFE_tensor_zeroquants}
\begin{eqnarray}
 && \nabla_{a}\nabla_{b}\Xi +\Xi L_{ab} - s g_{ab} =0 ,
  \label{StandardCEFEsecondderivativeCF}\\ &&
\nabla_{a}s +L_{ac} \nabla ^{c}\Xi=0 , \label{standardCEFEs}\\ &&
\nabla_{a}L_{bc}-\nabla_{b}L_{ac} - d^{d}{}_{cab}\nabla_d{}\Xi =0 ,
 \label{standardCEFESchouten}\\ && 
 \nabla_{e}d^{e}{}_{abc} =0 , \label{standardCEFErescaledWeyl}\\ &&
 R^c{}_{dab} - \Xi d^c{}_{dab} - 2( \delta^c{}_{[a}L_{b]d} -
 g_{d[a}L_{b]}{}^c) =0.
 \label{standardCEFERiemannDecomp}\\ &&
\lambda - 6 \Xi s + 3 \nabla_{a}\Xi \nabla^{a}\Xi =0
\end{eqnarray}
\end{subequations}
where $\Xi$ is the conformal factor, $L_{ab}$ is the Schouten tensor,
$d_{abcd}$ is the rescaled Weyl tensor and $s$ is the Friedrich
scalar, defined in terms of more familiar quantities via
\begin{align}\label{defs_conf_vars}
  s&:= \tfrac{1}{4}\nabla_{a}\nabla^{a}\Xi +
  \tfrac{1}{24}R\Xi,\\ L_{ab}&:=\tfrac{1}{2}R_{ab}-\tfrac{1}{12}Rg_{ab},
  \\ d^{a}{}_{bcd}&:=\Xi^{-1}C^{a}{}_{bcd}.
\end{align}
where $C^a{}_{bcd}$, $R^{a}{}_{bcd}$, $R_{ab}$ and $R$ are the Weyl,
Riemann, Ricci tensor and Ricci scalar of $\bmg$.

The CEFEs as written above can be thought as merely a set of
identities satisfied by the curvature tensors and other geometric
objects defined on $(\mathcal{M},\bmg)$.  The process of recasting the
standard Einstein field equations (or the CEFEs given above) as a set
of partial differential equations is called a hyperbolic reduction
since the equations are split into evolution and constraint equations,
being the former of hyperbolic type. The CEFEs are \emph{formally}
regular in the sense that there are no $\Xi^{-1}$ terms in equations
\eqref{CFE_tensor_zeroquants}. There are different hyperbolic
reduction strategies for the CEFEs and the one that
we are interested in, is the metric and second order hyperbolic
reduction of the CEFEs first introduced in \cite{Pae13} ---see 
\cite{CarHurVal} for the case where trace-free matter is included.
 An appealing property of this hyperbolic reduction is that
the metric satisfies a set of wave equations which looks formally
identical to the \emph{standard} Einstein field equations in
\emph{generalised harmonic gauge} ---which fixes the gauge through
(coordinate) gauge source functions $H^a$--- used in Numerical
Relativity, with the addition of a few extra terms.  These extra terms
are related to the fields $\Xi$, $s$, $L_{ab}$ and $d_{abcd}$ which
can be thought as some ``geometric-artificial-matter'' which, in turn,
satisfy another set of wave equations.  In fact, in this formulation
is convenient to use the trace-free Ricci tensor
$\Phi_{ab}:=\frac{1}{2}(R_{ab}-\frac{1}{4}Rg_{ab})$ instead of
$L_{ab}$ as en evolved variable. This is because in the CEFEs the
Ricci scalar $R$ is related to the choice of conformal representative
and hence is called the conformal gauge source function $F$. See
\cite{CarHurVal} and \cite{FenGas23} for further details.

Note that, despite their mathematically
appealing properties for the inclusion of~$\mathscr{I}$ in the numerical
domain, the constraint subsystem of the CEFEs is larger than in most common approaches.
Therefore, they pose an additional difficulty when one is to think about
numerical simulations of systems of physical interest, mainly because the constraints
will be unavoidably violated due to numerical error. Whether constraint violations
remain bounded in free evolution schemes will be studied elsewhere. In this
work we focus in the evolution sector of the linearised CEFEs in order to test the
scri-fixing technique of~\cite{FenGas23}, which is independent
of the initial data used for evolutions.

 
\subsection{ The linearised conformal Einstein field equations in second order form }
\label{sec:lin_wave_CEFE_gen}

In \cite{FenGas23} the linearisation of the second order formulation of the
CEFEs of \cite{CarHurVal} was presented. These equations were derived for any
background and they will only be presented schematically here.
At the end of this section we fix our attention to
conformally flat backgrounds and write the equations explicitly.

\medskip

Let $\bm\varphi$ encode all the geometric variables while $\bmf$
encodes the gauge quantities.  To linearise the second order metric
formulation of the CEFEs one starts by considering the split:
\begin{align}
  g_{ab}=\mathring{g}_{ab} + \delta g_{ab}, \qquad
  \Phi_{ab}= \mathring{\Phi}_{ab} + \delta \Phi_{ab},
  \qquad \Xi = \mathring{\Xi} + \delta \Xi, \qquad s= \mathring{s} +
  \delta s
\end{align}
where the ring-quantities represent an exact (background) solution to
the CEFEs and the delta-quantities are perturbations.  The term $\bmf$
 encodes the \emph{Lorenz gauge source functions} $F^a$
---related to the linearisation of the coordinate gauge source
functions $\delta H^a$--- and the perturbation of the conformal gauge
source function $\delta F$ ---see \cite{FenGas23} for further details.
With this notation, the linearised second
order CEFEs read
\begin{equation}\label{wave_linear_CEFEs}
  \begin{aligned}
    \mathring{\square} \delta g_{ab}
    &=
    H^{g}_{(ab)}(\delta \bm\varphi,
    \dotnabla\delta\bm\varphi \; ; \;  \mathring{\bm\varphi}, \bmf, 
    \dotnabla\mathring{\bm\varphi},
    \dotnabla \bmf )
    \\
    \mathring{\square} \delta  \Xi 
    & = H^{\Xi}_{}(\delta\bm\phi, \dotnabla \delta\bm\varphi
    \; ; \; 
    \mathring{\bm\varphi}, \bmf, \dotnabla\mathring{\bm\varphi} )  \\
    \mathring{\square} \delta  s 
    & = 
    H^{s}(\delta\bm\varphi, \dotnabla\delta\bm\varphi
    \; ; \; 
    \mathring{\bm\varphi}, \bmf, \dotnabla\mathring{\bm\varphi},
    \dotnabla \bmf )   \\ 
    \mathring{\square} \delta \Phi_{ab }
    &= 
    H^{\Phi}_{(ab)}(\delta \bm\varphi,
    \dotnabla \delta \bm\varphi \; ; \; 
    \mathring{\bm\varphi}, \bmf, \dotnabla\mathring{\bm\varphi},
      \dotnabla \bmf,  \dotnabla\dotnabla \mathring{F}, 
      \dotnabla\dotnabla \delta F )
    \\
    \mathring{\square} \delta  d_{abcd}
    & =
    H^{d}_{[ab][cd]}(\delta \bm\varphi, 
    \dotnabla \delta \bm\varphi
    \; ; \; 
    \mathring{\bm\varphi}, \bmf, \dotnabla\mathring{\bm\varphi},
    \dotnabla\dotnabla\mathring{\bm\varphi}, \dotnabla \bmf ) .
  \end{aligned}
\end{equation}
Here $\mathring{\square}:=\mathring{g}^{ab}\dotnabla_a\dotnabla_b$
where $\dotnabla$ denotes the Levi-Civita connection of
$\mathring{\bmg}$.  Observe that the semicolon separates the arguments
which do not contain the evolved perturbation variables. Notice that only the
equation for $\delta \Phi_{\mu\nu}$ contains second derivatives of
$\delta F$ which restricts the allowed choices for $\delta F$.

\section{The background unphysical spacetime and scri-fixing}

In this section the geometry of the conformal background is discussed
as well as the scri-fixing strategy via gauge source functions for the
linearised CEFEs of~\cite{FenGas23}.

\subsection{The inversion-Minkowski spacetime}

%
 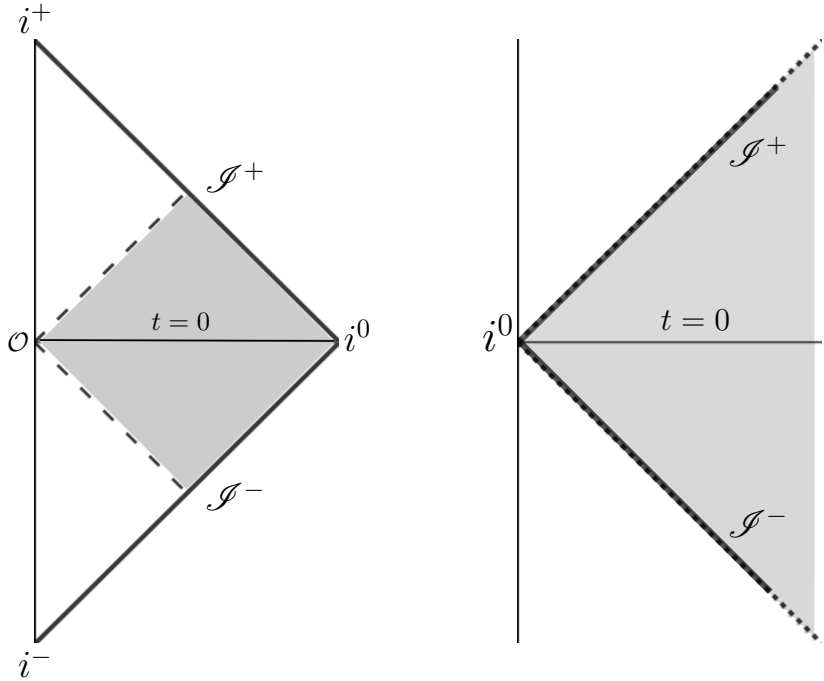
\begin{figure}[t!]
  \begin{center}
     \begin{subfigure}{}
   \includegraphics[width=4cm]{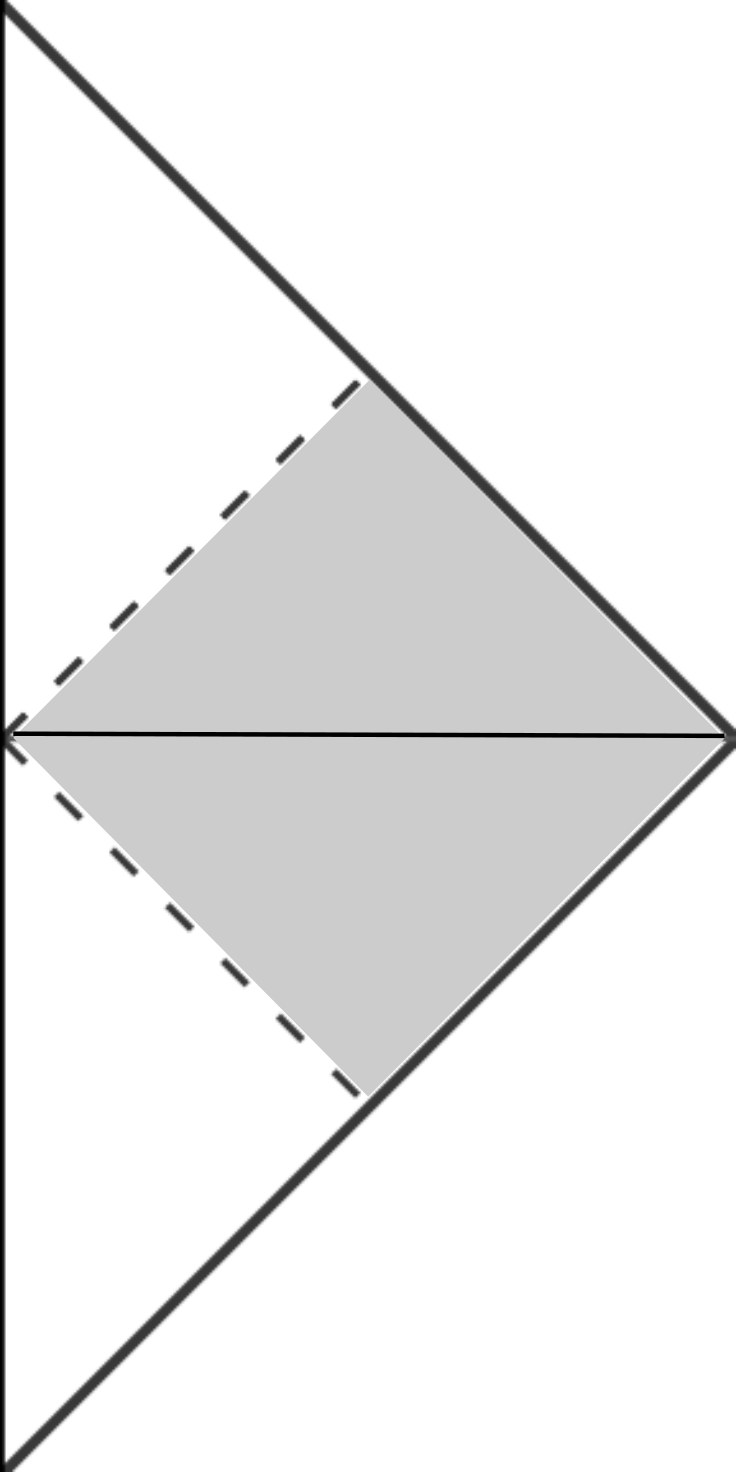}
   \put(-120,230){\Large{$i^+$}}
   \put(-50,170){\Large{$\mathscr{I}^{+}$}} \put(-70,118){$t=0$}
   \put(2,110){\Large{$i^0$}} \put(-124,110){\Large{$\smallO$}}
   \put(-50,50){\Large{$\mathscr{I}^{-}$}}
   \put(-120,-13){\Large{$i^-$}}
     \end{subfigure}
     \hspace{2cm}
     \begin{subfigure}{}
     \includegraphics[width=4cm]{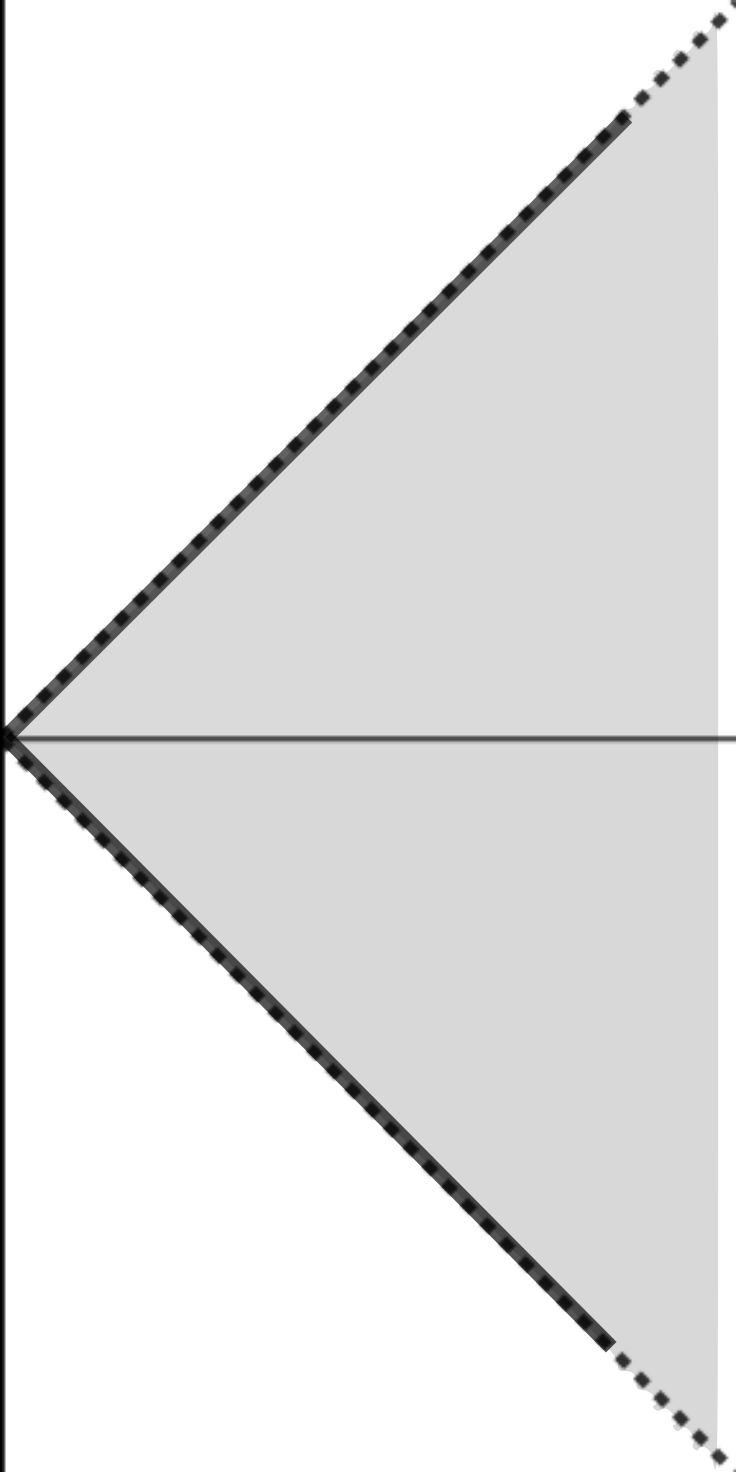}
     \put(-35,40){\Large{$\mathscr{I}^{-}$}}
     \put(-35,180){\Large{$\mathscr{I}^{+}$}}
     \put(-127,110){\LARGE{$i^0$}} \put(-60,118){\large{$t=0$}}
     \end{subfigure}
   \end{center}
  \caption{The left panel shows the Penrose diagram of the Minkowski
    spacetime where the shaded region corresponds to that covered by the \emph{inversion
    conformal representation}.  As depicted in the right panel, in the
    inversion-Minkowski spacetime, $\mathscr{I}^{\pm}$ is represented by the
    lightcone through the unphysical origin which, in turn, corresponds to
    $i^0$.   }\label{fig:diagram}
 \end{figure}
  

The (physical)
Minkowski spacetime metric in spherical polar coordinates reads
\begin{align}\label{MinkowskiMetricPhysicalPolar}
\tilde{\bmeta}=-\mathbf{d}\tilde{t}\otimes\mathbf{d}\tilde{t}
+\mathbf{d}\tilde{r}\otimes \mathbf{d}\tilde{r}+\tilde{r}^2
\mathbf{\bm\sigma},
\end{align}
with~$\tilde{t}\in(-\infty, \infty)$, $\tilde{r}\in [0,\infty)$
  where~$\bm\sigma$ denotes the standard metric on~$\mathbb{S}^2$.
  Recall that for any conformally flat metric $\bmg= \Xi^2\tilde{\bm\eta}$
  in general one has:
  \begin{eqnarray}
\Phi_{ab}  \neq 0, \qquad R  \neq 0, \qquad  d_{abcd}=0.
  \end{eqnarray}
  However, there exist a non-trivial rescaling ($\Xi \neq const$) for which
  the unphysical metric $\bmg$ is again flat. This is a classical conformal
  representation of Minkowski that can be consulted in \cite{Ste91}.
  To see this explicitly, one introduces unphysical coordinates $(t,r)$
  given by:
\begin{align}
 \label{physicalToUnphysicaltrho}
   t=\frac{\tilde{t}}{\tilde{r}^2-\tilde{t}^2}, \qquad r =
   \frac{\tilde{r}}{\tilde{r}^2-\tilde{t}^2}.
\end{align}
A direct calculation then shows that
\begin{align}
  \bmeta = \Xi^2 \tilde{\bmeta}
\end{align}
where
\begin{align}
\label{InverseMinkowskiUnphysicaltrhocoords}
\bm\eta=-\mathbf{d}t\otimes\mathbf{d}t +\mathbf{d}r\otimes
\mathbf{d}r+r^2 \mathbf{\bm\sigma}, \qquad \Xi=r^2 -t^2.
\end{align}

 One can verify that the associated Friedrich scalar is constant
 $s=2$.  We call this conformal representation the
 \emph{inversion-Minkowski} spacetime since the roles of spatial
 infinity $i^0$ and the physical origin $\smallO$ are swapped. Hence,
 in the unphysical picture future and past null infinity
 $\mathscr{I}^\pm$ correspond to the lightcone through the unphysical
 origin.  Therefore, the inversion-Minkowski spacetime is not a full
 compactification of the physical spacetime in the sense that
 $\smallO$ is located at $r\rightarrow \infty$ and the future and past
 timelike infinities $i^\pm$ are also not included in the
 domain. Despite this drawback notice that one can still have access
 to a portion of $\mathscr{I}^{\pm}$ ---see diagram \ref{fig:diagram}.
 Naturally, one could opt to choose another conformal compactification
 of the Minkowski spacetime such as the one that conformally embeds
 the Minkowski spacetime into the Einstein cylinder (used for instance
 to obtain the textbook Penrose diagrams of Minkowski, de Sitter and
 anti-de Sitter spacetimes) so that one has a full compactification of
 the spacetime.  This is of particular interest, but comes with the
 price that $\Phi_{ab} \neq 0$ and $R \neq 0$, which renders most of
 the linearised equations \eqref{wave_linear_CEFEs} (except that for
 the equation for the perturbation of the Weyl tensor) excessively
 long and cumbersome to deal with. Hence, the inversion-Minkowski
 spacetime is chosen for its simplicity and clean implementation but
 from the formulation point of view one could have equally opted for
 an analogous implementation in the Einstein's cylinder background.

\subsection{The linearised CEFEs around a flat background}
\label{sec:flat}

In this section we write the linearised second order metric CEFEs
for the particular case in
which $\mathring{\bmg}$ is flat, this achieved non-trivially
only in the case of the inversion conformal representation
of the Minkowski spacetime as discussed in the previous section.
In this case the background quantities read
\begin{eqnarray}
  \mathring{\Phi}_{ab}  = \mathring{R} = \mathring{d}_{abcd}=0,\qquad
  \mathring{\Xi}=r^2-t^2, \qquad \mathring{s}=2.
\end{eqnarray}
This is the case for which the general expression simplify the most and they explicitly read:

 \begin{subequations}\label{LinCEFEprop}
  \begin{align}
    \mathring{\square}\delta g_{\mu \nu } & = -4 \delta \Phi _{\mu \nu
    } - \mathring{g}_{\mu \nu } \dotnabla _{\alpha }F^{\alpha } +
    2\dotnabla _{(\mu }F_{\nu ) } + \tfrac{1}{2}\mathcal{F}
    \dotg_{\mu\nu},
    \label{linMetricClean}
    \\ \mathring{\square} \delta \Xi & = 4 \delta s - \mathring{s}
    \delta g ^{\mu }{}_{\mu } + F^{\mu}\dotnabla_\mu \mathring{\Xi} -
    \tfrac{1}{6}\dotXi \mathcal{F},
  \label{linCF}\\
   \mathring{\square} \delta s & = - \tfrac{1}{6}\mathring{s}
   \mathcal{F} - \tfrac{1}{6}\dotnabla_\mu\mathcal{ F} \dotnabla^\mu
   \dotXi,
  \label{linFscalar} \\
  \mathring{\square} \delta \Phi_{\mu \nu } & =
  \tfrac{1}{6}\dotnabla_{\{\mu}\dotnabla_{\nu\}} \mathcal{ F},
  \label{linTFRicci} \\
    \mathring{\square} \delta d_{\mu\nu\alpha\beta}
    &=0. \label{linResWeyl}
  \end{align}
 \end{subequations}
 here parenthesis represent symmetrisation while curly brackets denote
 taking the symmetric trace-free part. Also, to simplify the notation
 we have denoted $\delta F$ by $\mathcal{F}$ so that only perturbation
 evolved quantities have a $\delta$ decoration.

 \subsection{Gauge sources and scri-fixing strategies}\label{sec:genscrifixingstrategy}

The linear analogue of the conformal and coordinate gauge source
functions are encoded via $\mathcal{F}$ and $F^a$, respectively.  An
advantageous property of the generalised harmonic gauge formulation is
that one can leverage the choice of gauge source functions.
For instance in hyperboloidal formulation of
\cite{GasHil18, PetGauVanHil24} the gauge source functions are chosen
to drive the asymptotic behaviour of the fields. In contrast, in the
case of the CEFEs current numerical and analytical studies make use of
the so-called conformal Gaussian gauge \cite{FraSte21, FraSte23,
  FraGooSte23, GasVal17}. Although the conformal Gaussian gauge is a
very geometrical choice for which the conformal factor is not longer
an unknown it comes with a price: the transformation connecting the
``physical'' and ``unphysical'' coordinates is not known explicitly
but rather fixed through a set of ordinary differential equations (the
conformal geodesic equations) which more often than not, cannot be
solved in closed form.  Even in the cases where these expressions can
be reduced to quadratures it is not possible to write, in an simple
and closed form way, physically important metrics such as that of the
Schwarzschild or Kerr spacetimes ---see for instance \cite{GarGasVal18,
  Fri03a}.  This is in stark contrast with the hyperboloidal approach
of \cite{Hil16, VanHusHil14} where the change of coordinates from
physical to unphysical coordinates and/or conformal factor is given in
a well-motivated but essentially ad-hoc way. This is particularly
advantageous as this allows to write most physically relevant
spacetimes in hyperboloidal coordinates in an explicit and closed form
---see for instance \cite{PanMac24}.  In \cite{FenGas23} it was
noticed that, at least for the linear case, one can leverage the gauge
source functions to achieve a scri-fixing similar to that of the
hyperboloidal approach but for the linearised CEFEs.  To do so,
observe that one can allow $F^a$ to be a function of the coordinates
as well as the fields $\bm\varphi$ while $\mathcal{F}$ can depend on
the coordinates (but not $\bm\varphi$) without breaking the
hyperbolicity of these equations. Also notice that the zeros of the
conformal factor $\Xi$ do not necessarily coincide as those of the
background conformal factor $\mathring{\Xi}$ as one has $\Xi=
\mathring{\Xi} + \delta \Xi$. Hence, scri-fixing for the linear case
simply means being able to ensure that $\delta \Xi=0$ so that the
zeros of the full conformal factor $\Xi$ correspond to those of that
of the background conformal factor $\mathring{\Xi}$.  In the current
linear set up one can achieve this scri-fixing by simply choosing
$F^a(x;\bm\varphi)$ such that

\begin{eqnarray}\label{Scri-fixingGauge}
F^{a} \dotnabla _{a}\dotXi =   \tfrac{1}{6} \dotXi \mathcal{F} - 4 \delta s + \dotXi \dotPhi ^{ab}
  \delta g_{ab}   + \mathring{s} \delta g^{a}{}_{a}.
\end{eqnarray}

Therefore $\mathring{\square}\delta \Xi=0$. Hence, setting trivial
initial data for $\delta \Xi$ in conjunction with existence and
uniqueness of wave equations will ensure that $\delta\Xi=0$. Notice
however that, this method already shows its limitations regarding the
problem at $i^0$ since in order to solve for $F^a$ in equation
\eqref{Scri-fixingGauge} one needs $\dotnabla_{a}\dotXi \neq 0$. In
other words, this gauge choice works for any slicing for which
$\dotnabla_a \mathring{\Xi} \neq 0$ everywhere. A particular example
of these slices are hyperboloidal slices as they precisely avoid the
neighbourhood of $i^0$. The scri-fixing method for the linearised
CEFEs of \cite{FenGas23} is very similar in spirit to the scri-fixing
technique introduced for the hyperboloidal approach
in \cite{Zen07a, Zen08}. Conceptually, the difference is that method
is to be applied to a formally regular set of equations (CEFEs)
instead of a formally singular one. It should be
pointed out, however, that there exist already in the CEFEs literature
a similar choice called the scri-freezing shift gauge \cite{Fra98a}.
Understanding the relation between these three gauge choices aimed to
control the location of $\mathscr{I}$ will be explored elsewhere.

\section{Numerical implementation and results}
In this section we describe the numerical implementation of the equations and
the results of the numerical evolutions performed for this work.

\subsection{3+1 decomposition}\label{sec:3p1Decomp}

Although equations \eqref{LinCEFEprop} are clearly hyperbolic ---as
long as $\mathcal{F} = \mathcal{F} (x^\mu) $ and
$F^a=F^a(\bm\phi,x^\mu)$--- for numerical purposes it is convenient to
perform a 3+1 split of both the evolved fields and the background
geometry.  The customary 3+1 decomposition will not be repeated here
and this section will simply serve to set up the notation and
conventions for the subsequent discussion.  The background spacetime
metric is decomposed as $\mathring{g}_{ab}=
\mathring{\gamma}_{ab}-\mathring{n}_a\mathring{n}_b$, where
$\mathring{\bmn}$ is the normal to some fiduciary spacelike
hypersurface $\mathcal{S}$ with induced metric $\bm\gamma$.  The
extrinsic curvature $\mathring{\bmK}$, normal $\mathring{\bmn}$ and
acceleration vector $\mathring{\bma}$ are related via
$\mathring{\nabla}_a\mathring{n}_b=\mathring{K}_{ab}+\mathring{a}_b\mathring{n}_a$
where
$\mathring{a}^a=-\mathring{n}^b\mathring{\nabla}_b\mathring{n}^a$.

\noindent 
The tensorial perturbation variables are decomposed accordingly:
 \begin{align*}
   && \delta g_{nn}:=\mathring{n}^a\mathring{n}^b\delta g_{ab}, &&
   \delta g_{na}:=\mathring{\gamma}_{a}{}^c\mathring{n}^b\delta
   g_{cb}, && \delta
   \gamma_{ab}:=\mathring{\gamma}_{a}{}^c\mathring{\gamma}_{b}{}^d
   \delta g_{cd}.\\ && \delta
   \Phi_{nn}:=\mathring{n}^a\mathring{n}^b\delta \Phi_{ab}, && \delta
   \Phi_{na}:=\mathring{\gamma}_{a}{}^c\mathring{n}^b\delta \Phi_{cb},
   && \delta
   \phi_{ab}:=\mathring{\gamma}_{a}{}^c\mathring{\gamma}_{b}{}^d
   \delta \Phi_{cd}\\ && \delta E_{ab}:=
   \mathring{n}^b\mathring{n}^d\delta d_{abcd}, && \delta
   B_{ab}:=\mathring{n}^b\mathring{n}^d\delta d_{abcd}^{\star}.
\end{align*}
 where $\delta d_{abcd}^{\star} :=
 \frac{1}{2}\mathring{\epsilon}_{cd}{}^{pq}\delta_{abpq}$ and
 $\mathring{\epsilon}_{abcd}$ is the volume form of $\mathring{\bmg}$.  Finally,
 the Lorenz source functions are decomposed as
 \begin{eqnarray}
F_n=\mathring{n}^aF_a, && f_a=\mathring{\gamma}_{a}{}^cF_c.
 \end{eqnarray}
 Since the numerical infrastructure we will use (NRPy+
 \cite{RucEtiBau18}, see the next subsection for more details) is
 built for first order in time and second order in space equations, in
 the final equations to be implemented we have introduced
 time-reduction-variables defined through the Lie derivative along
 $\mathring{\bmn}$. For instance we define the time reduction variable
 $\delta \phi_{ab}^{\pi}:= \mathcal{L}_{\mathring{\bmn}} \delta
 \phi_{ab}$ and then use the wave equation \eqref{linTFRicci} to
 obtain a first order in time evolution equation for
 $\delta\phi_{ab}^{\pi}$.  We proceed similarly with all the
 variables. The final form of the equations will not be displayed
 here. This calculation was carried out using \texttt{xAct} in
 \texttt{Mathematica}.
 Additionally, using equation \eqref{Scri-fixingGauge} one obtains that, in this
 case, the gauge source functions that achieve the scri-fixing are
 those satisfying:
 
 \begin{align}\label{eq:Scri-Fix-GaugeSources-MinkInverted}
   - t F_{n}{} + r f^{r} = -2 \delta s + \tfrac{1}{12} \mathcal{F} r^2
   - \tfrac{1}{2}\mathring{s}\delta g_{nn}{} - \tfrac{1}{12}
   - \mathcal{F} t^2 \tfrac{1}{2} \mathring{s}\mathring{\gamma}^{ij}
   - \delta\gamma _{ij}.
 \end{align}
 
  For the numerical evolutions discussed in the next section two sets
  of gauge source functions will be used, one for which all the gauge
  source functions are zero and other for which all vanish except for
  $F_n$, which satisfies equation
  \eqref{eq:Scri-Fix-GaugeSources-MinkInverted}.

   \begin{figure}[t!]
  \begin{center}
     \begin{subfigure}{}
   \includegraphics[width=4cm]{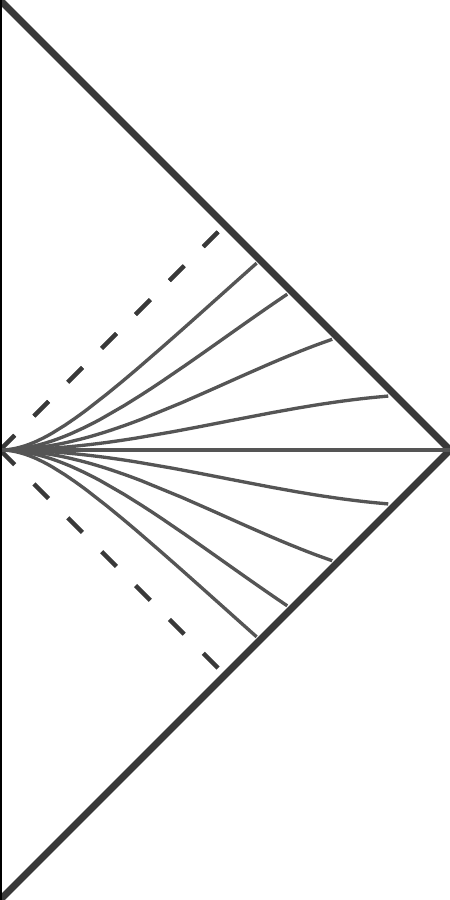}
   \put(-120,230){\Large{$i^+$}}
   \put(-50,170){\Large{$\mathscr{I}^{+}$}}
   \put(2,110){\Large{$i^0$}} \put(-124,110){\Large{$\smallO$}}
   \put(-50,50){\Large{$\mathscr{I}^{-}$}}
   \put(-120,-13){\Large{$i^-$}}
     \end{subfigure}
     \hspace{2cm}
     \begin{subfigure}{}
     \includegraphics[width=4cm]{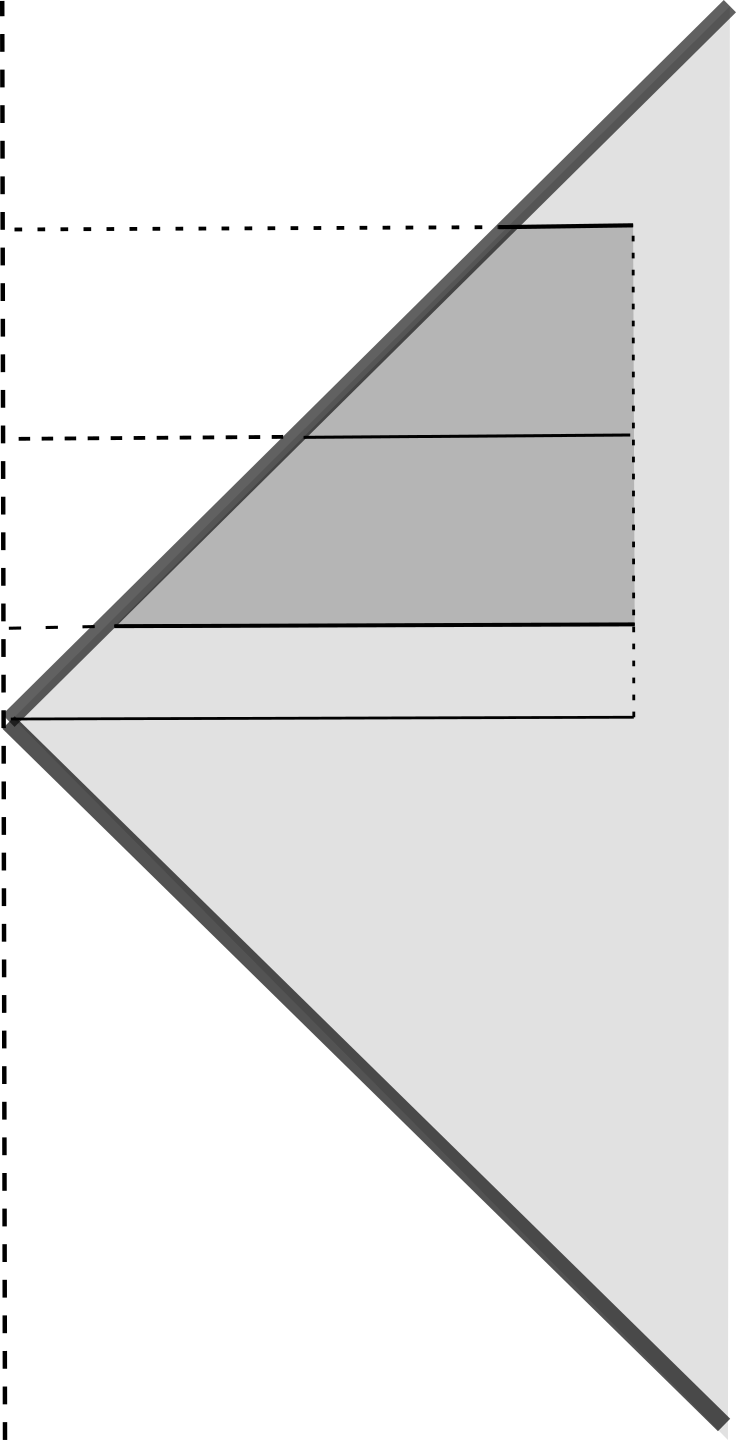}
     \put(-50,10){\Large{$\mathscr{I}^{-}$}}
     \put(-50,200){\Large{$\mathscr{I}^{+}$}}
     \put(-127,110){\LARGE{$i^0$}} \put(-10,123){\large{$t=t_{i}$}}
     \put(-10,183){\large{$t=t_{f}$}}
     \put(-20,102){\large{$r_\bullet$}}
     \end{subfigure}
   \end{center}
  \caption{The left panel shows the $t=const$ surfaces plotted on the
    Penrose diagram of the Minkowski spacetime.  The right panel shows
    a coordinate diagram of the inversion-Minkowski spacetime where the
    $t=const$ surfaces are shown. The numerical evolutions are started
    at $t=t_i\neq 0$ to avoid $i^0$ and the grid corresponds to
    $r\in(0,r_\bullet]$.}\label{fig:diagram3}
   \end{figure}
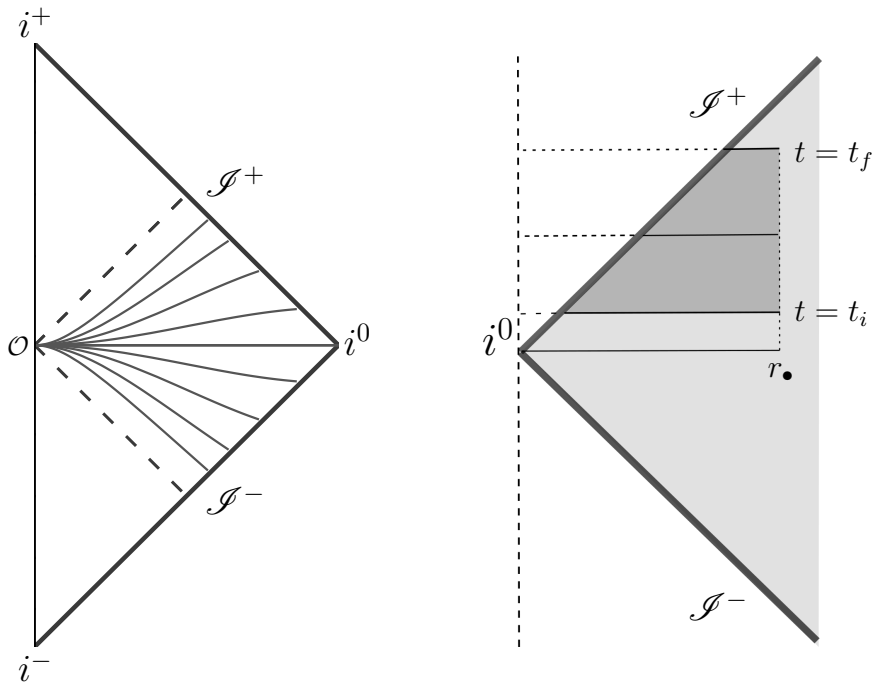

\subsection{Numerical evolutions}

We have implemented the 3+1 decomposed equations described in the
previous subsection in coordinates~$(t,r,\theta,\phi)$ using the NRPy+
software \cite{RucEtiBau18}. NRPy+ is a python infrastructure that
allows the input of tensorial expressions in Einstein notation and
outputs highly optimised C code that compiles and runs within the same
infrastructure. Hence, NRpy+ is well suited for the numerical
evolutions performed here, as the system of interest comprises a total
of 68 evolved variables in three spatial dimensions, several of which
satisfy long differential equations. Despite these challenges, we are
able to perform the numerical evolutions in standard desktop computers
thanks to the simplicity and efficiency of NRPy+.  The code uses the
method of lines with a fourth-order Runge-Kutta for time
integration. Note that for the inversion-Minkowski spacetime
background the slice~$t=0$ intersects~$i^0$, where the scri-fixing
technique discussed in this paper does not work, as mentioned in
subsection \ref{sec:genscrifixingstrategy}, so we are forced to start our
simulation at some later time. We choose~$t=0.5$ as the starting time
for all our evolutions ---see diagram \ref{fig:diagram3}.  Moreover,
since the inversion-Minkowski spatial coordinates~$(r,\theta,\phi)$
mimic standard spherical coordinates in physical Minkowski spacetime,
we discretise the spatial dimensions employing a staggered grid with
equally spaced~$(N_r,N_\theta,N_\phi)$ cells in the~$(r,\theta,\phi)$
directions and one gridpoint lying exactly at the centre of each cell.
This way we avoid the direct evaluation of quantities at the
coordinate singularities, corresponding to the~$z$ axis.  The spatial
domain consists of~$r\in (0,40)$, $\theta \in [0,\pi]$ and
$\phi\in[0,2\pi]$.  Spatial derivatives are approximated with
second-order accurate centred finite differences.

\medskip

Notice that background quantities appear in the evolution equations,
as clearly shown in equations~\eqref{LinCEFEprop}.  In particular, for
the inversion-Minkowski spacetime the background conformal factor
$\mathring{\Xi}$ is a time-dependent quantity.  For different
spacetimes one could consider, several background quantities can have
this property. One relevant extension of the present
work is the implementation of time-dependent right-hand-sides for the
evolved quantities within the NRPy+ infrastructure, which was not
available in the first public version of the code. We expect to be able
to exploit this tool in future work.  As previously mentioned, the
inversion-Minkowski spacetime background was chosen because of its
simplicity, however, it should be emphasised that the (background) location
of~$\mathscr{I}^+$ is determined by~$\mathring{\Xi}=0$, which
corresponds to the points such that $r=t$ (see the given expression in
\eqref{InverseMinkowskiUnphysicaltrhocoords}). Hence, as our grid is
fixed in the~$(r,\theta,\phi)$ coordinates, this implies that as time
evolves we have increasing number of gridpoints lying outside of the
corresponding physical Minkowski spacetime ---see diagram
\ref{fig:diagram3}.  This problem is commonly faced in the numerical
evolution of the CEFEs, as generically the conformal factor depends on
time.

\medskip

  One way to remedy loosing larger parts of the grid into the
  ``unphysical region'' of the conformal extension, would be to use
   a choice of hyperboloidal foliation according to the
    prescription in \cite{Zen07a} for which the background conformal
  factor is not time dependent. Despite the fact that the scri-fixing
  through gauge source functions for the linearised CEFEs discussed in section
  \ref{sec:genscrifixingstrategy} can accommodate for this possibility,
  the price to pay is that the background curvature becomes
  non-zero. This makes the equations more cumbersome to deal with and
  the background spatial metric is no longer the flat metric that is
  already incorporated in NRPy+. Although these  modifications
  do not represent a problem at the level of the formulation,
  incorporating more general backgrounds would involve making further
  changes to the NRPy+ code.  Hence, we have opted to stick to the
  inversion-Minkowski background for the proof-of-concept numerical
  evolution presented here and leave other more interesting
  backgrounds such as the hyperboloidal representation of Minkowski of
  \cite{Zen07a} or the conformal representation of Minkowski,
  de-Sitter and anti-de Sitter in the Einstein cylinder for future
  work.

\medskip
 
As is standard in NRPy+, we rescale the angular components of the
tensorial quantities by the appropriate factor to evolve purely
$O(1)$ quantities near the points~$r\sin\theta=0$. This implies that
for every covariant $\theta$ index we rescale this component by a
factor of $r$ and for every $\phi$ index with a factor
$r\sin\theta$. More concretely, if $\Lambda_\theta$ is any tensorial
component appearing in the equations of motion we write it as
$\Lambda_\theta = r\tilde{\Lambda}_\theta$ and evolve
$\tilde{\Lambda}_\theta$. On the contrary, contravariant angular
components such as say $\tilde{\Lambda}^\theta$ and
$\tilde{\Lambda}^\phi $ are rescaled by $r^{-1}$ and
$(r\sin\theta)^{-1}$ respectively.  These rescalings provide a better
setup to deal with the coordinate singularities, which lie closer to
the gridpoints as resolution is increased.  We also apply artificial
dissipation to all evolved variables to improve convergence and
stability at the origin. We do this via fourth order Kreiss-Oliger
dissipation with a dissipation parameter~$\sigma=0.02$. Importantly,
we increase resolution near the coordinate singularities with a factor
in front of the Kreiss-Oliger angular derivatives of tensorial
quantities according to the prescription described in the previous
paragraph, namely, a~$(r\sin\theta)^{-2}$ in front
of~$\partial_\phi^2$ of a rank-2 tensor, etc.

\medskip

There are two types of numerical boundaries, the so-called
\textit{inner} and \textit{outer}. The inner ones correspond to the
boundaries of the angular coordinates or the points where~$r<0$. To
fill ghost points beyond the boundaries of the angular coordinates we
simply input the corresponding value from the gridpoint inside the
grid. For the~$r<0$ gridpoints we do so with the appropriate parity
conditions that must be satisfied by the different tensorial
components. All of this is standard in NRPy+ and is explained in
\cite{RucEtiBau18}. For the outer boundary we use fourth order
extrapolation in all the variables. Note there is no physical
motivation for both boundary conditions in $r$, as the points~$r<0$
are outside of the physical Minkowski spacetime and~$r>40$ corresponds
to the finite value of $\tilde{r}$ for which there could be
incoming radiation. These facts imply that our numerical solution only
represents physically viable scenarios in the future domain of
dependence of our initial data. A deeper study on how to include the
the whole physical domain ---i.e containing the physical origin
$\tilde{r}=0$--- with appropriate boundary conditions will also be
addressed in an upcoming work.

\newpage

\begin{figure*}[h!]
  \begin{center}
 \includegraphics[scale=0.6]{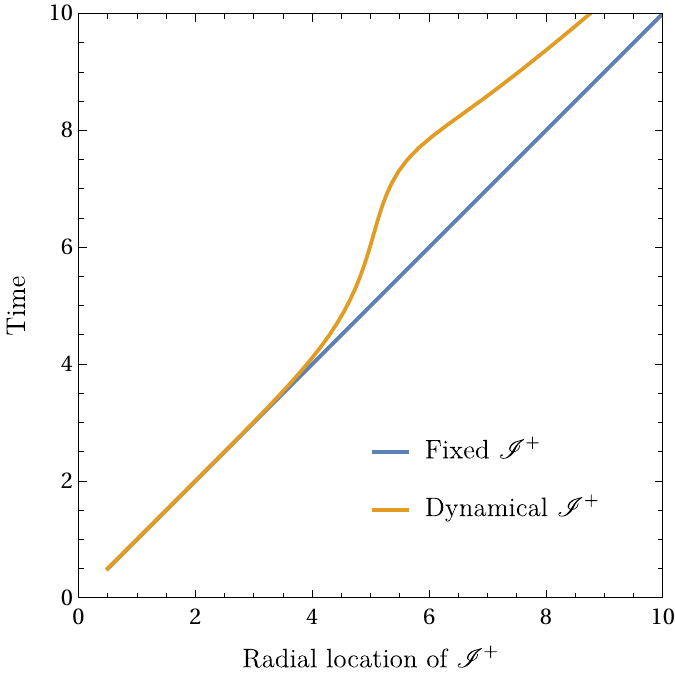}
 \caption{Location of~$\mathscr{I}^+$ as a function of time for both
   scri-fixing and trivial gauge source functions.}
 \label{scri-location_sphsymm}
 \end{center}
\end{figure*}

Finally, a comment about initial data is in order. If a solution to
equations~\eqref{LinCEFEprop} is to represent a solution to the
linearised CEFEs we need to solve the linearised version of the
CEFE-constraints to determine the initial data. This requires a
thorough study on how to numerically solve the constraints, which
lies out of the scope of the present work. Ultimately, for this
proof-of-concept experiment we are concerned with the numerical
evolution with the scri-fixing gauge sources, which is independent of
the initial data solution. Therefore we test all of our numerical
simulations with constraint-violating initial data.

\subsubsection{Spherically symmetric evolutions}

As a first test, we perform numerical evolutions with spherically
symmetric initial data, which implies that the whole evolution remains
spherically symmetric. We
run our simulations with~$N_r=160$ and~$N_\theta =
N_\phi = 2$ gridpoints, both for scri-fixing and trivial gauge source
functions. Setting two gridpoints in the coordinates that are 
not being used is a technical requirement of NRPy+.


\noindent As initial data we set for all the evolved fields:
\begin{equation*}
 S_{\text{sph}}(r) = A_s r^n e^{ -(r^2 - c^2)^2 / 4\sigma^2 },
\end{equation*}
where~$c = 10$ for all variables,~$A_s=0$ for $\delta \Xi(0,r)$,
$A_s=10^{-2}$ and $n=0$ for~$\delta g_{nn}(0,r) = \delta s(0,r)$,
$A_s=10^{-2}$ and $n=2$ for~$\delta\Phi_{nn}(0,r)=\delta \Phi_{rr}(0,r)
= \delta E_{rr}(0,r) = \delta B_{rr}(0,r)$ and $A_s=10^{-2}$ and $n=3$
for~$\delta \Phi_{nr}(0,r) = \delta g_{nr}(0,r)$. Notice that
the condition $\delta\Phi_{nn}(0,r)=\delta \Phi_{rr}(0,r)$ ensures that
$\delta\Phi_{ab}$ is trace-free. Finally we set
$A_s=-10^{-2}/2$ and $n=4$ for $\delta E_{\theta\theta} =
\delta E_{\phi\phi}/\sin^2\theta = \delta B_{\theta\theta} =
\delta B_{\phi\phi}/\sin^2\theta $ to guarantee~$\delta E_{ij}$ and $\delta B_{ij}$ are
trace-free. We set zero initial data for the remaining components and
for all reduction variables.

\medskip

\noindent 
In figure~\ref{scri-location_sphsymm} we plot the location
of~$\mathscr{I}^+$ as a function of time for our two sets of gauge
source functions. In the scri-fixing case, this plot coincides with
the~$r=t$ line as $\delta \Xi=0$, whereas the other case does not, as
expected. The curve representing the location of~$\mathscr{I}^+$ in
the trivial gauge sources case (labelled as ``dynamical'' in
figure~\ref{scri-location_sphsymm}) is found through a root-finding method at
each timestep. Hence, it can, in general, become non-smooth even for
smooth initial data if~$\delta\Xi$ becomes non-monotonic as a function
of $r$.

\begin{minipage}{\textwidth}
 \hspace{-0.1cm}\includegraphics[scale=0.6]{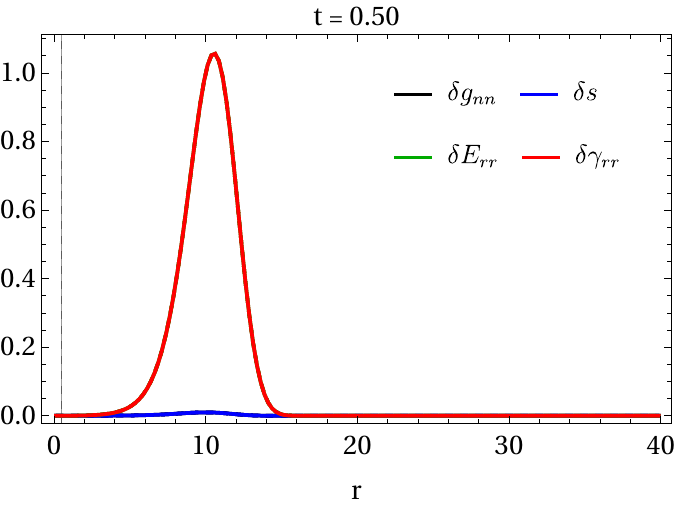}
 \hspace{-0.0cm}\includegraphics[scale=0.6]{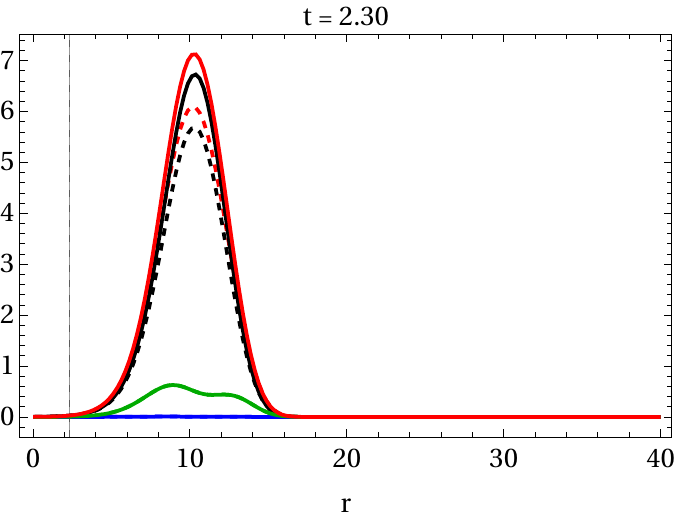}
 \hspace{-1.2cm}\includegraphics[scale=0.6]{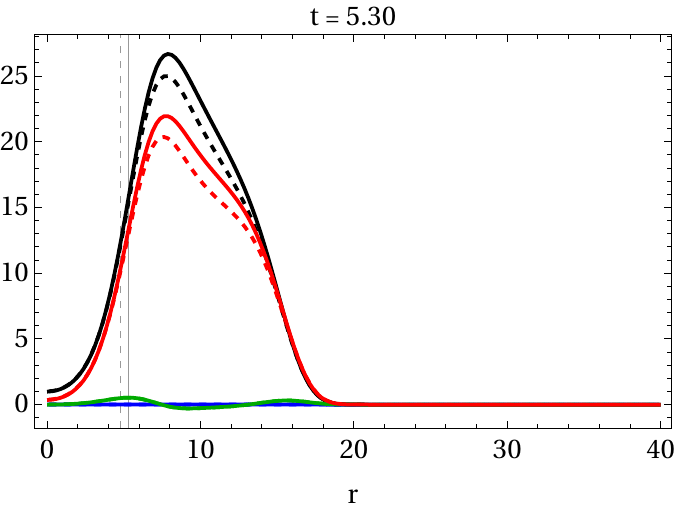}
 \hspace{1cm}\includegraphics[scale=0.6]{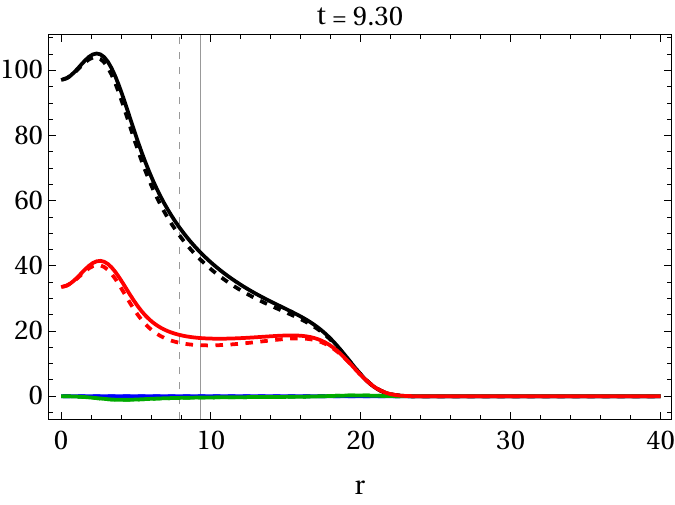}
 \captionof{figure}{Snapshots of evolved fields for both scri-fixing
   and trivial gauge source functions.}
 \label{sphsymm_snapshots}
\end{minipage}

\begin{figure*}[h!]
\centering
\includegraphics[scale=0.6]{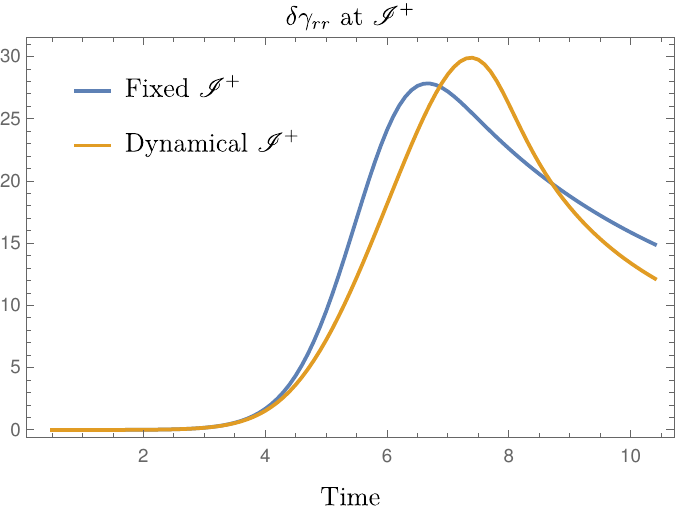}
\includegraphics[scale=0.6]{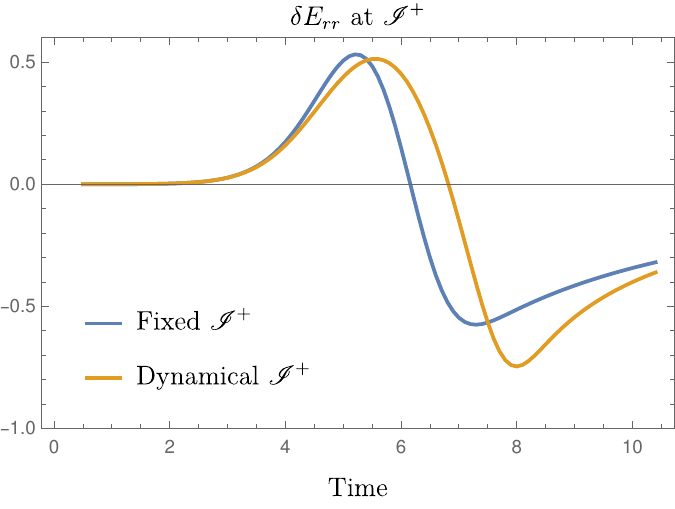}
 \caption{Example of evolved fields evaluated at the points plotted in
   figure~\ref{scri-location_sphsymm} for both sets of gauge sources.}
 \label{fields-atscri_sphsymm}
\end{figure*}

\medskip
\noindent In figure~\ref{sphsymm_snapshots} we plot snapshots of some
of the evolved fields both with scri-fixing gauge sources (solid
lines) and trivial gauge sources (dashed lines). Additionally, we plot
the location of $\mathscr{I}^+$ at each time with a vertical line,
also with a solid or dashed line depending on each case. Note that the
location of~$\mathscr{I}^+$ at each time is found by solving the
equation~$\mathring{\Xi}+\delta\Xi = 0$. Since~$\delta\Xi = 0$ for
scri-fixing (up to numerical error) and~$\delta\Xi\neq 0$ for trivial
gauge sources, the location of $\mathscr{I}^+$ differs between these
two cases in the same way as shown in figure~\ref{scri-location_sphsymm}
above. Moreover, since the scri-fixing gauge sources enter the
right-hand-side of evolution equations of the the metric
perturbations, both~$\delta g_{nn}$ and~$\delta\gamma_{rr}$ differ as
functions of code coordinates during the evolution.  On the other
hand, coordinate gauge sources do not appear in the equations of
motion for~$\delta s$,~$\delta E_{ij}$ or~$\delta B_{ij}$ and they are
decoupled from the rest. Therefore, these fields coincide in the
evolution in the code's coordinates for the two sets of gauge sources
we test, as shown in figure~\ref{sphsymm_snapshots}.  Finally, having
found the location of~$\mathscr{I}^+$ as a function of time we can
evaluate the fields directly at those values by interpolating the
numerically evolved fields. Examples with the two sets of gauge
sources are plotted in figure~\ref{fields-atscri_sphsymm}.  Note that,
despite~$\delta E_{ij}$, $\delta B_{ij}$ and~$\delta s$ being
identical as functions of the coordinates, their evaluation
at~$\mathscr{I}^+$ differs between the two sets of gauge sources
because the location of~$\mathscr{I}^+$ is different.  This
``ambiguity'' ultimately comes the decoupling of the Bianchi-sector
from the metric-sector of the linearised equations and from
interpreting the location of $\mathscr{I}^{+}$ as either the locus
points of $\mathring{\Xi}=0$ or $\Xi:=\mathring{\Xi}+\delta \Xi
=0$. In the non-linear case there is no such decoupling and hence no
ambiguity. Therefore, the scri-fixing gauge sources in the linear case
can be thought as a way to remove this ambiguity by ensuring that
$\delta \Xi=0$.

\subsubsection{Axially symmetric evolutions}

\begin{figure}[h!]
 \centering
 \includegraphics[scale=0.28]{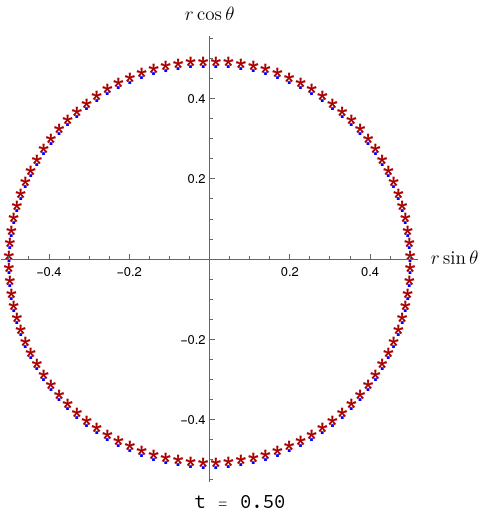}
 \includegraphics[scale=0.28]{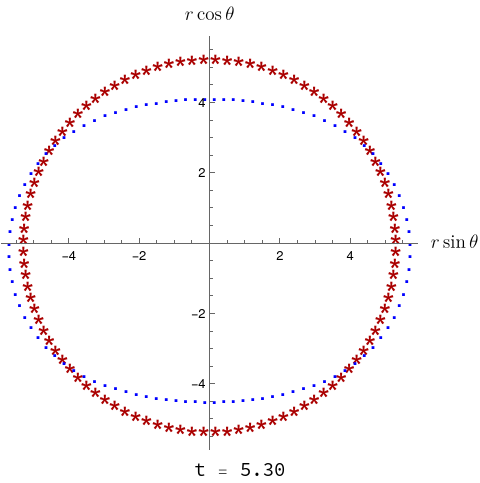}
 \includegraphics[scale=0.28]{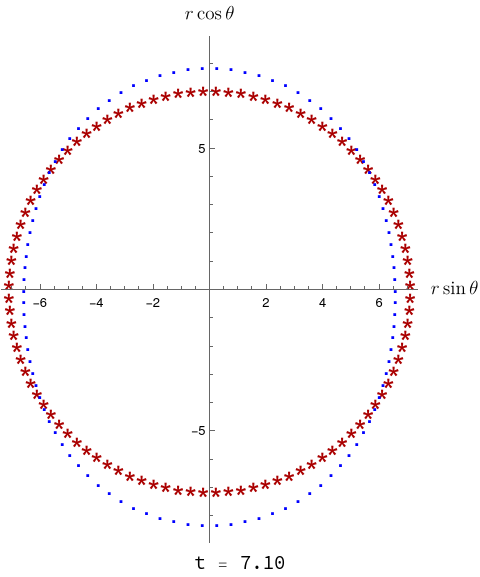}
 \includegraphics[scale=0.28]{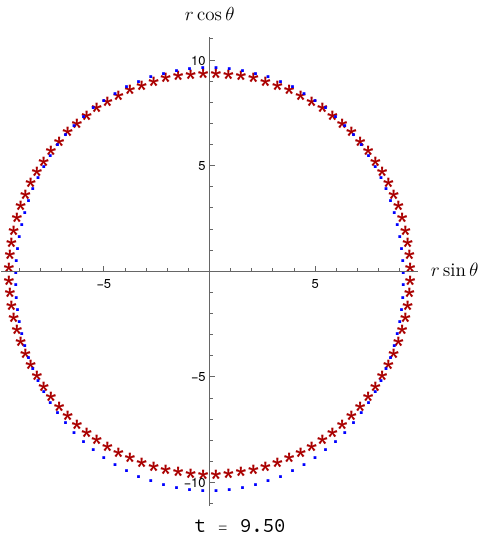}
 \caption{Snapshots of the location of $\mathscr{I}^{+}$.}
 \label{scri-location_axsymm}
\end{figure}

We then proceed with simulations in axial symmetry,
for both sets of gauge sources. As our first simulation we take $N_r =
80$, $N_\theta = 50$ and $N_\phi = 2$ gridpoints. Similarly to the
spherically-symmetric case we set as initial data for all variables
\begin{eqnarray*}
 S_{\text{ax}}(r) &=& A_a r^n R(r) Y_{20}(\theta,\phi) \\ R(r) &=&
 \left[ (4c^2r^2 +8cr^3 +4r^4 -2c\sigma^2r -4\sigma^2r^2
   -\sigma^4)/\sigma^4 \right. \\ & & \left. \left( -4c^2r^2 +8cr^3
   -4r^4 -2c\sigma^2r +4\sigma^2r^2 +\sigma^4
   \right)e^{4cr/\sigma^2}/\sigma^4 \right] \frac{e^{-(r+c)^2/\sigma^2
 }}{r}
\end{eqnarray*}
where~$Y_{20}$ is the~$l=2$, $m=0$ spherical harmonic and $c=8$ in this
case. This initial data is smooth at the origin, with the expression
inspired by the d’Alembert partial-wave solution to the wave
equation \cite{FerVicHil21}. We make identical choices for~$n$
and~$A_a=A_s/2$ for exactly the same fields as in the spherically symmetric case.
In figure~\ref{scri-location_axsymm} we plot snapshots of the location
of~$\mathscr{I}^+$ in the~$yz$ plane. Note that, once again, in the
scri-fixing case the plots coincides with the~$r=t$ curve, drawing a
circle in our coordinates, and in the trivial gauge sources case they
are deformed, as expected.

In Figures~\ref{gammarr_axsnapshots} - \ref{Err_axsnapshots} we plot
snapshots of~$\delta\gamma_{rr}$ and~$\delta E_{rr}$.  In these
figures, the first row corresponds to evolutions with trivial gauge
sources while the second row correspond to the case of scri-fixing
gauge sources. The colours blue and orange in these plots represent
the physical and unphysical region of the conformal extension respectively.
The location of~$\mathscr{I}^+$ corresponds to the boundary between
these two, which is again found by solving the
equation~$\mathring{\Xi}+\delta\Xi = 0$ for every angle.
Analogous to the spherically-symmetric case, the evolution of~$\delta
\gamma_{rr}$ as function of code coordinates differs between the
different sets of gauge sources, contrary to the evolution of~$\delta
E_{rr}$, which is unaffected. This feature can be seen comparing first
and second rows, respectively. For the chosen data, the difference is,
nevertheless, small.  If one considers a larger amplitude $A_a$ for the
initial data the difference is much more noticeable, however the
routine used to find the location of $\mathscr{I}$ has difficulties to
find the roots giving non-smooth features in Figure
\ref{scri-location_axsymm}.

\begin{minipage}[t]{\textwidth}
  \vspace{5mm}
 \includegraphics[scale=0.28]{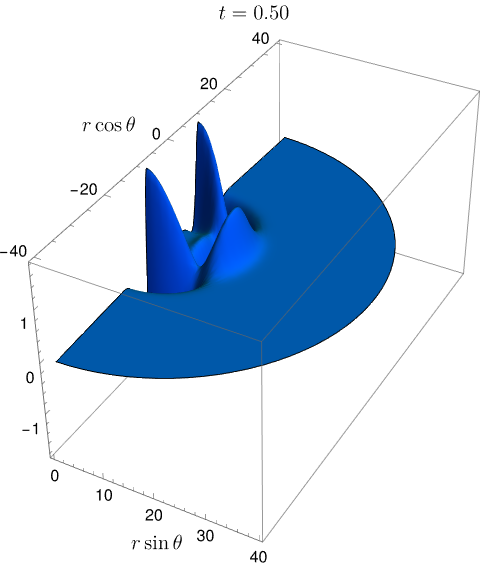}
 \includegraphics[scale=0.28]{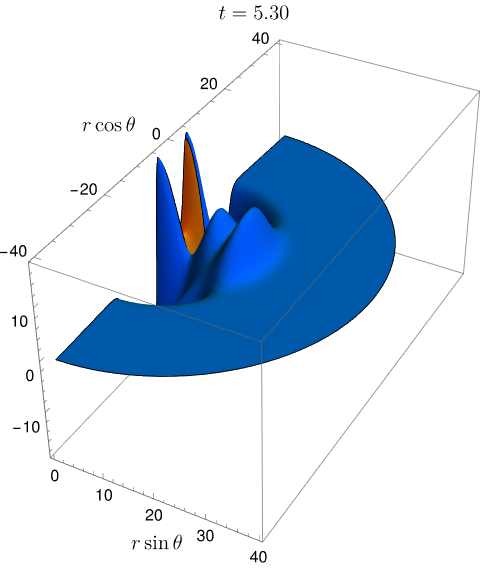}
 \includegraphics[scale=0.28]{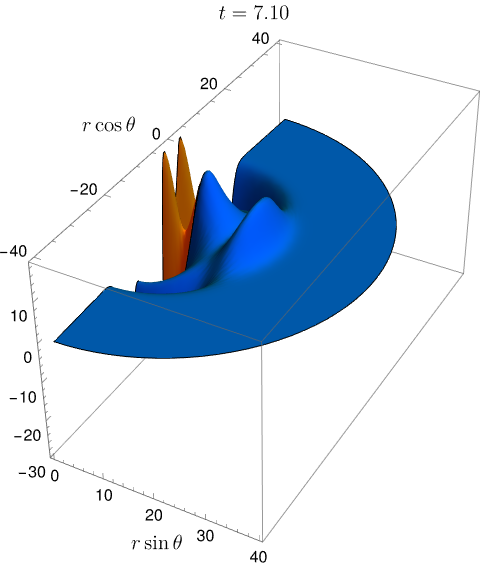}
 \includegraphics[scale=0.28]{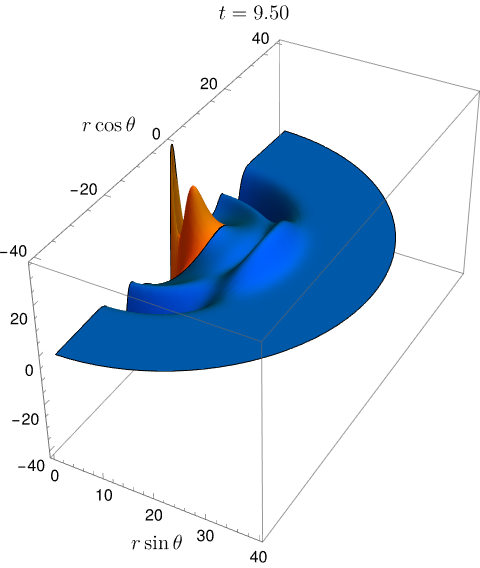}
 
 \includegraphics[scale=0.28]{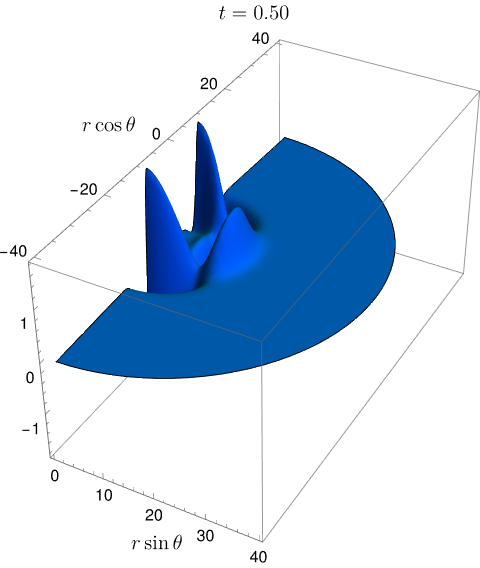}
 \includegraphics[scale=0.28]{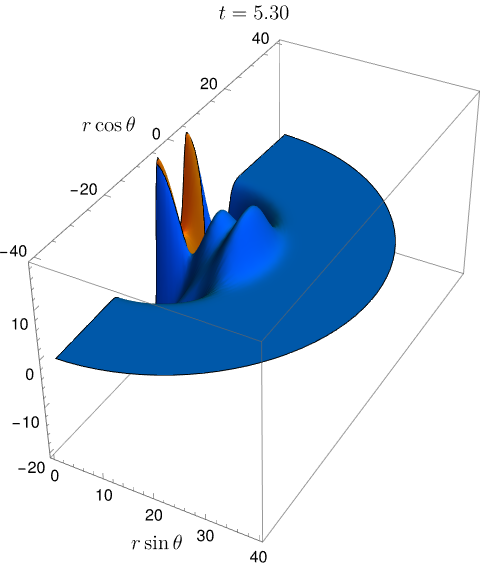}
 \includegraphics[scale=0.28]{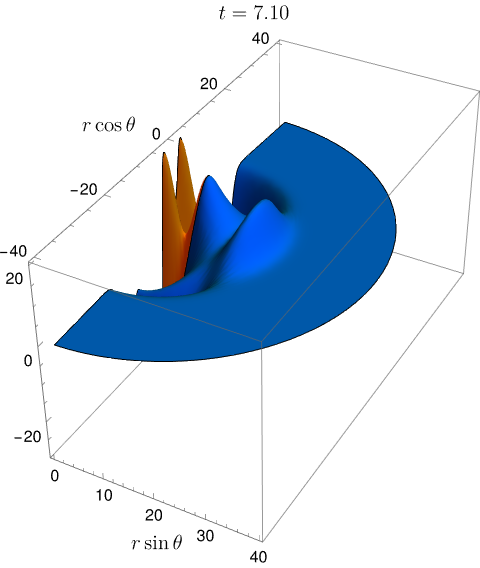}
 \includegraphics[scale=0.28]{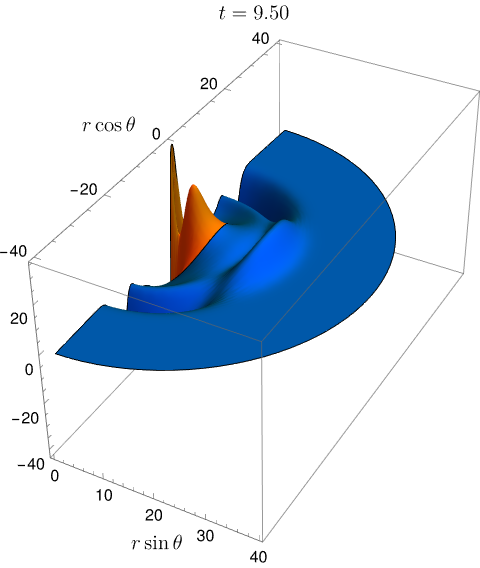}
 \captionof{figure}{$\delta \gamma_{rr}$ snapshots with trivial gauge
   sources (top) and scri-fixing gauge sources (bottom).}
 \label{gammarr_axsnapshots}
\end{minipage}

\begin{minipage}{\textwidth}
 \includegraphics[scale=0.28]{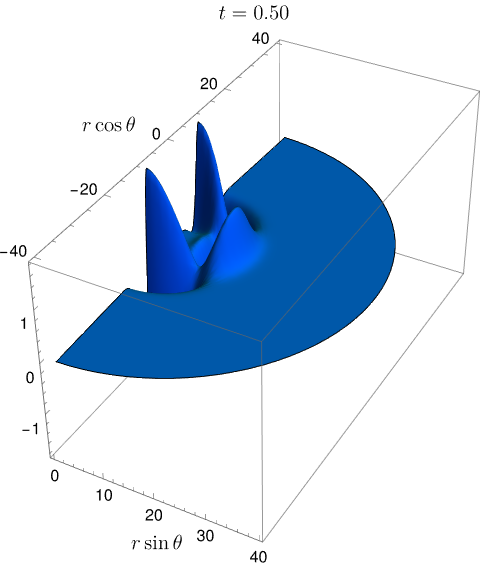}
 \includegraphics[scale=0.28]{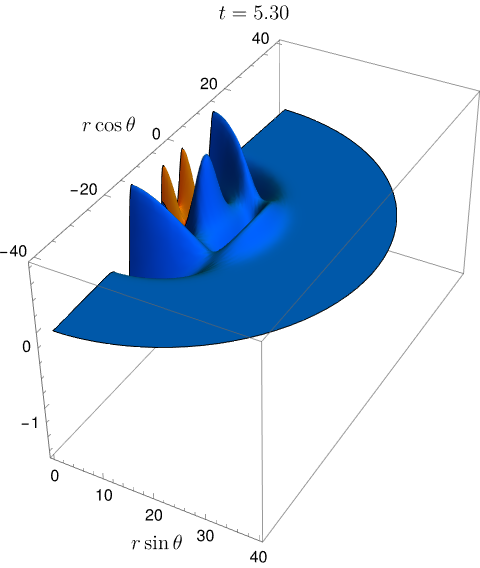}
 \includegraphics[scale=0.28]{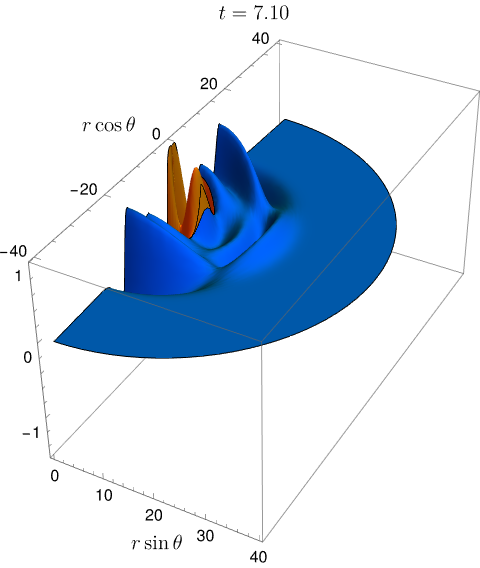}
 \includegraphics[scale=0.28]{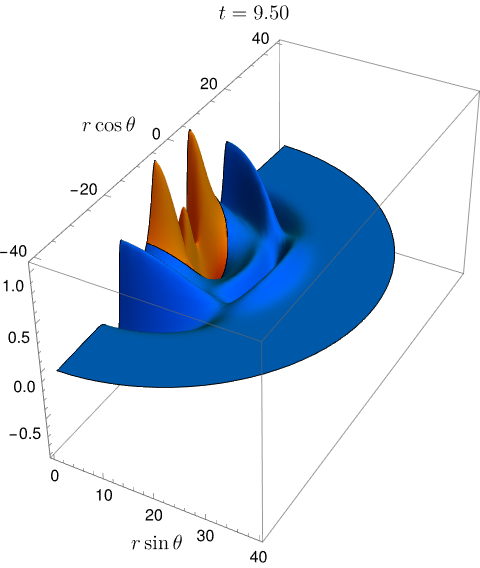}
 
 \includegraphics[scale=0.28]{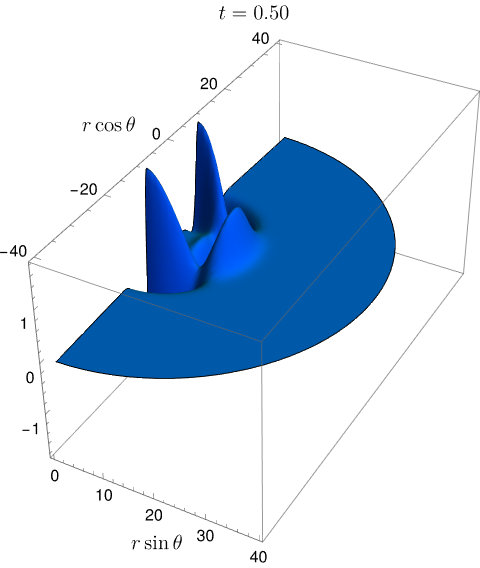}
 \includegraphics[scale=0.28]{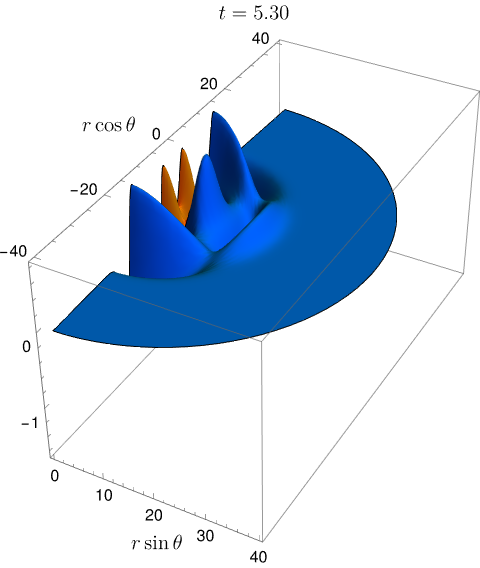}
 \includegraphics[scale=0.28]{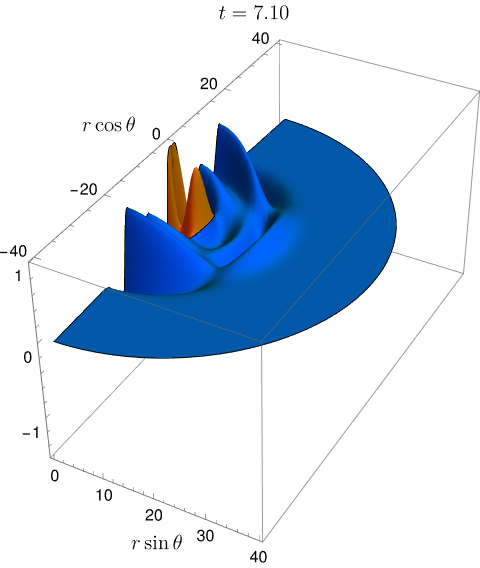}
 \includegraphics[scale=0.28]{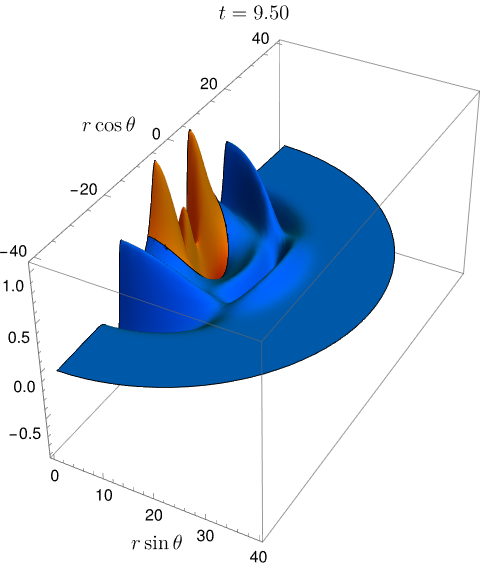}
 \captionof{figure}{$\delta E_{rr}$ snapshots with trivial gauge
   sources (top) and scri-fixing gauge sources (bottom).}
 \label{Err_axsnapshots}
\end{minipage}

\medskip

\begin{figure}[t!]
 \centering
 \includegraphics[scale=0.4]{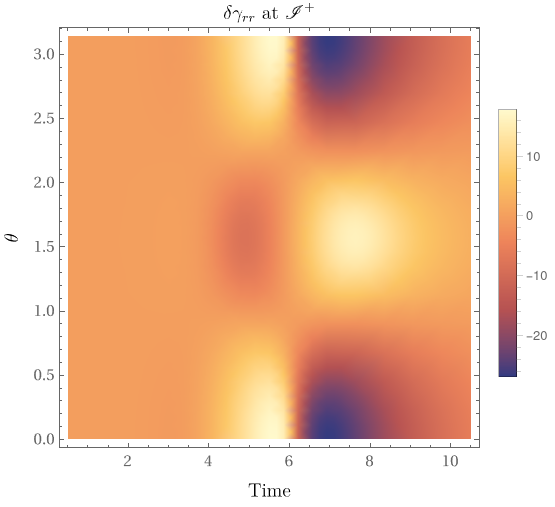}
 \includegraphics[scale=0.4]{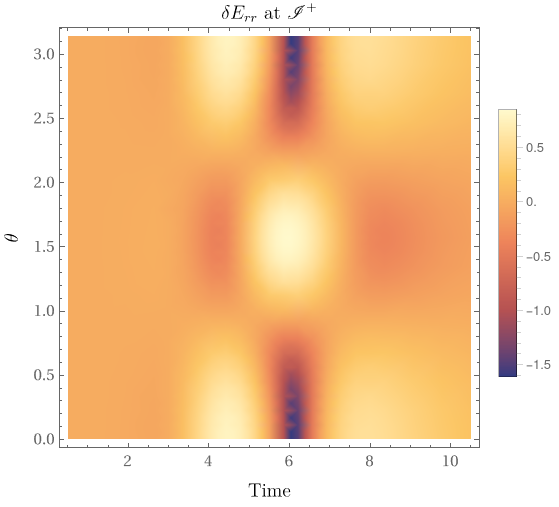}
 \includegraphics[scale=0.4]{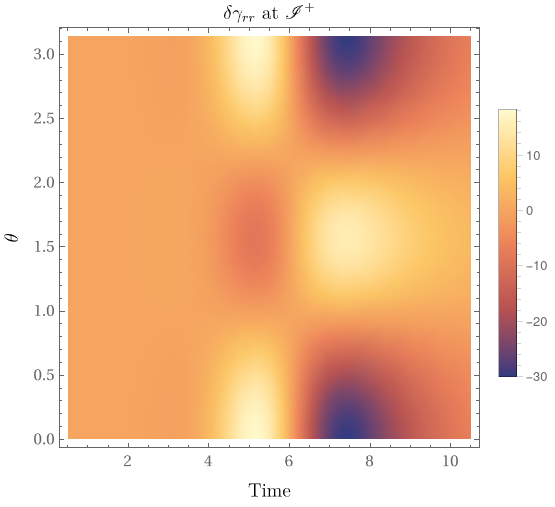}
 \includegraphics[scale=0.4]{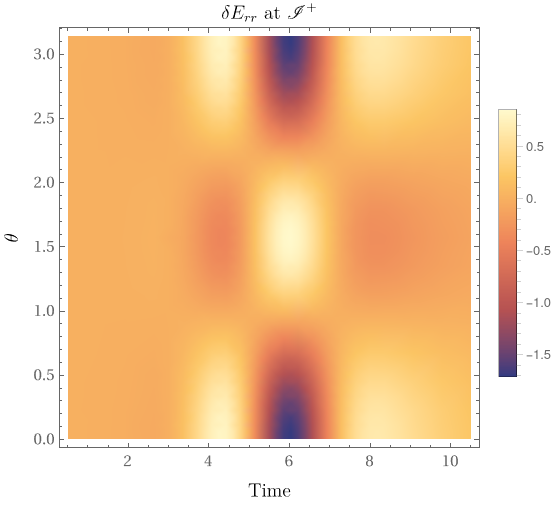}
 \caption{Trivial gauge sources (top) and scri-fixing gauge sources
   (bottom).}
 \label{fieldsatscri_axsymm}
\end{figure}

\begin{figure}[h!]
\centering
\includegraphics[scale=0.8]{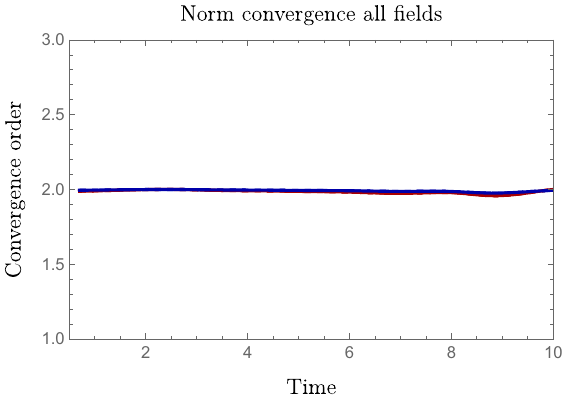}
\caption{Norm convergence for both cases. The red line corresponds to
  the case of trivial gauge source functions and the blue line to
  scri-fixing gauge source functions.}
 \label{convergence_test}
\end{figure}

With the location of~$\mathscr{I}^+$ in terms of our coordinates we
are able to extract the value of the evolved fields in these
points. In figure~\ref{fieldsatscri_axsymm} we plot again~$\delta \gamma_{rr}$
and~$\delta E_{rr}$ to illustrate this. The fields now depend
also on the angular coordinate~$\theta$, so we denote the amplitude of
the fields in colour. Note that, despite~$\delta E_{rr}$ being the same
function of~$(t,r)$ for both sets of gauge sources, its value
at~$\mathscr{I}^+$ differs between the two cases because of the
difference in the location of~$\mathscr{I}^+$, as discussed
previously.
We then proceed to do a self-convergence test of our simulations by
tripling $N_r$ and $N_\theta$ twice, starting with $N_r=80$ and~$N_\theta=16$
and computing the difference between adjacent resolutions in the overlapping gridpoints.
This is done for both sets of gauge sources. We then compute the~$L^2$ norm of these two
differences of all evolution quantities and compute the corresponding convergence order according to the formula
\begin{equation}
\textrm{Convergence order} = \log_3\left(\frac{\sqrt{\sum_{i=1}^{N_r}\sum_{j=1}^{N_\theta}(X_{low, i, j}-X_{med, i, j})^2}}{\sqrt{\sum_{i=1}^{N_r}\sum_{j=1}^{N_\theta}(X_{med, i, j}-X_{high, i, j})^2}}\right)
\end{equation}
with $X_{low/med/high, i, j}$ denoting each of the evolution variables for the low,
medium and high resolutions at point $i,j$\,.
In figure~\ref{convergence_test} we plot the convergence order as a function of time.
Note that, since we use second-order finite
differences, ideal convergence would correspond to the number 2. This
is compatible with figure~\ref{convergence_test}, meaning the
numerical errors are converging away at the expected rate in the limit
of infinite resolution.

    \section{Conclusions and perspectives}

    In this article we perform numerical evolutions of a linearised
    version of the second order metric formulation of the CEFEs introduced in
    \cite{Pae13} ---see also \cite{CarHurVal}---, which has the
    advantage that it resembles more closely the standard formulation of
    the Einstein field equations used in Numerical Relativity than
    other hyperbolic reductions of the CEFEs.
    Although the construction of initial data is not addressed, the
    numerical evolutions performed here serve as a proof-of-concept
    calculation to illustrate numerically the scri-fixing strategy
    through gauge source functions for the linear case discussed in
    \cite{FenGas23}. The main advantage of this scri-fixing strategy
    is that it allows to choose the conformal factor in an essentially
    ad-hoc way without solving any differential equation.  If this strategy
    is indeed generalisable to the non-linear case then it
    would allow to employ in the CEFEs the same gauges
    (same conformal factor) used in the formally singular hyperboloidal
    approach \cite{Van23, PetGauVanHil24} but with formally regular
    equations.  To test these ideas at the linear level, we have used
    the linear equations of \cite{FenGas23} in the simplest possible
    background: the inversion-Minkowski spacetime.  Notice that
    similar evolutions can be performed in other backgrounds at the
    expense of more complicated expressions. In particular, one could
    use the Einstein cylinder to perform evolutions of the Minkowski,
    de-Sitter and anti-de Sitter spacetimes.  Naturally, in all these
    cases the equation for the perturbation of the rescaled Weyl
    tensor decouples from the rest. This observation has been
    exploited in other explorations of the spin-2 equation
    representing the linearisation of the Bianchi sector of the
    CEFEs \cite{BeyDouFraWha13, MacVal18, Val03a}.  However, the
    philosophy in this article is that the complete system is to be
    evolved as relevant physical information is also encoded in the metric
    perturbation.

    \section*{Acknowledgements}

    We are grateful to Zach Etienne and Leonardo Rosa Werneck for
    answering questions regarding the implementation of
    time-dependent equations in NRPy. We have
    benefited from conversations with David Hilditch.  C. Peterson
    thanks the financial support to IST-ID through Project No. UIDB/00099/2020
    E. Gasper\'in holds an FCT (Portugal)
    investigator grant 2020.03845.CEECIND and held an FCT Exploratory
    Research Project 2022.01390.PTD during the time this research was
    conducted.
    A. Va\~n\'o Vi\~nuales thanks FCT for financial support through Project~No.~UIDB/00099/2020
    and for the FCT funding with DOI 10.54499/DL57/2016/CP1384/CT0090.




\begin{thebibliography}{10}

\bibitem{FenGas23}
Justin Feng and Edgar Gasperín.
\newblock Linearised conformal einstein field equations.
\newblock {\em Classical and Quantum Gravity}, 40(17):175001, July 2023.

\bibitem{HilNur16}
C~Denson Hill and Pawel Nurowski.
\newblock How the green light was given for gravitational wave search, 2016.

\bibitem{BieGarYun17}
Lydia Bieri, David Garfinkle, and Nicolas Yunes.
\newblock Gravitational waves and their mathematics, 2017.

\bibitem{Bon62}
H.~Bondi, M.~G.~J. van~der Burg, and A.~W.~K. Metzner.
\newblock Gravitational waves in general relativity. vii. waves from
  axi-symmetric isolated systems.
\newblock {\em Proceedings of the Royal Society of London. Series A,
  Mathematical and Physical Sciences}, 269(1336):21--52, 1962.

\bibitem{Sac62}
R.~K.~: Sachs and Hermann Bondi.
\newblock Gravitational waves in general relativity viii. waves in
  asymptotically flat space-time.
\newblock {\em Proceedings of the Royal Society of London. Series A.
  Mathematical and Physical Sciences}, 270(1340):103--126, 1962.

\bibitem{NewPen62}
Ezra Newman and Roger Penrose.
\newblock An approach to gravitational radiation by a method of spin
  coefficients.
\newblock {\em Journal of Mathematical Physics}, 3(3):566--578, 1962.

\bibitem{Ger77}
Robert~P. Geroch.
\newblock Asymptotic structure of space-time.
\newblock In F.~Paul Esposito and Louis Witten, editors, {\em Asymptotic
  Structure of Space-Time}, pages 1--105, Plenum Press, New York, 1977.
  Springer US.

\bibitem{Ash81}
Abhay Ashtekar.
\newblock Radiative degrees of freedom of the gravitational field in exact
  general relativity.
\newblock {\em Journal of Mathematical Physics}, 22(12):2885--2895, 12 1981.

\bibitem{Fer24}
Francisco Fernández-Álvarez.
\newblock News tensor on null hypersurfaces, 2024.

\bibitem{Ash14}
A.~{Ashtekar}.
\newblock {Geometry and Physics of Null Infinity}.
\newblock {\em ArXiv e-prints}, September 2014.

\bibitem{Pen64}
Roger Penrose.
\newblock Conformal treatment of infinity.
\newblock In C.~DeWitt and B.~DeWitt, editors, {\em Relativity, Groups, and
  Topology (Les Houches, France, 1964)}, pages 565--584. Gordon and Breach, New
  York, 1964.

\bibitem{Zen08}
Anil Zenginoglu.
\newblock {Hyperboloidal evolution with the Einstein equations}.
\newblock {\em Class. Quant. Grav.}, 25:195025, 2008.

\bibitem{Van15}
Alex Va{\~n}{\'o}-Vi{\~n}uales.
\newblock {\em {Free evolution of the hyperboloidal initial value problem in
  spherical symmetry}}.
\newblock PhD thesis, U. Iles Balears, Palma, 2015.

\bibitem{Van23}
Alex Vañó-Viñuales.
\newblock Spherically symmetric black hole spacetimes on hyperboloidal slices.
\newblock {\em Frontiers in Applied Mathematics and Statistics}, 9, August
  2023.

\bibitem{HilHarBug16}
David Hilditch, Enno Harms, Marcus Bugner, Hannes R{\"u}ter, and Bernd
  Br{\"u}gmann.
\newblock {The evolution of hyperboloidal data with the dual foliation
  formalism: Mathematical analysis and wave equation tests}.
\newblock {\em Class. Quant. Grav.}, 35(5):055003, 2018.

\bibitem{PetGauVanHil24}
Christian Peterson, Shalabh Gautam, Alex Va\~n\'o Vi\~nuales, and David
  Hilditch.
\newblock {Spherical evolution of the generalized harmonic gauge formulation of
  general relativity on compactified hyperboloidal slices}, 2024.

\bibitem{Fri81}
H.~{Friedrich}.
\newblock {On the Regular and the Asymptotic Characteristic Initial Value
  Problem for Einstein's Vacuum Field Equations}.
\newblock {\em Proc. R. Soc. Lond. A}, 375(1761):169--184, March 1981.

\bibitem{Val16}
Juan-Antonio Valiente-Kroon.
\newblock {\em Conformal Methods in General Relativity}.
\newblock Cambridge University Press, Cambridge, 2016.

\bibitem{Fri02}
Helmut Friedrich.
\newblock Spin-2 fields on minkowski space near spacelike and null infinity.
\newblock {\em Classical and Quantum Gravity}, 20(1):101--117, dec 2002.

\bibitem{GasVal17}
Edgar Gasper{\'i}n and Juan~A. Valiente~Kroon.
\newblock Perturbations of the asymptotic region of the schwarzschild--de
  sitter spacetime.
\newblock {\em Annales Henri Poincar{\'e}}, pages 1--73, 2017.

\bibitem{MinVal23}
Marica Minucci and Juan~A Valiente~Kroon.
\newblock On the non-linear stability of the cosmological region of the
  schwarzschild-de sitter spacetime.
\newblock {\em Classical and Quantum Gravity}, 40(14):145005, June 2023.

\bibitem{DouFra16}
Georgios Doulis and J\"org Frauendiener.
\newblock Global simulations of minkowski spacetime including spacelike
  infinity.
\newblock {\em Phys. Rev. D}, 95:024035, Jan 2017.

\bibitem{Hub99}
Peter H{\"u}bner.
\newblock A scheme to numerically evolve data for the conformal {E}instein
  equation.
\newblock {\em Class. Quantum Grav.}, 16:2823--2843, 1999.

\bibitem{Hub01}
Peter H{\"u}bner.
\newblock From now to timelike infinity on a finite grid.
\newblock {\em Class. Quantum Grav.}, 18:1871--1884, 2001.

\bibitem{FraSte21}
J~Frauendiener and C~Stevens.
\newblock The non-linear perturbation of a black hole by gravitational waves.
  i. the bondi–sachs mass loss.
\newblock {\em Classical and Quantum Gravity}, 38(19):194002, September 2021.

\bibitem{FraGooSte23}
Jörg Frauendiener, Alex Goodenbour, and Chris Stevens.
\newblock The non-linear perturbation of a black hole by gravitational waves.
  iii. newman-penrose constants, 2023.

\bibitem{AnsMac16}
Marcus Ansorg and Rodrigo Panosso~Macedo.
\newblock {Spectral decomposition of black-hole perturbations on hyperboloidal
  slices}.
\newblock {\em Phys. Rev.}, D93(12):124016, 2016.

\bibitem{ZenNunHus09}
Anıl Zenginoğlu, Darío Núñez, and Sascha Husa.
\newblock Gravitational perturbations of schwarzschild spacetime at null
  infinity and the hyperboloidal initial value problem.
\newblock {\em Classical and Quantum Gravity}, 26(3):035009, January 2009.

\bibitem{JarMacShe21}
Jos\'e~Luis Jaramillo, Rodrigo Panosso~Macedo, and Lamis Al~Sheikh.
\newblock {Pseudospectrum and Black Hole Quasinormal Mode Instability}.
\newblock {\em Phys. Rev. X}, 11(3):031003, 2021.

\bibitem{MacZen24}
Rodrigo~Panosso Macedo and Anil Zenginoglu.
\newblock Hyperboloidal approach to quasinormal modes, 2024.

\bibitem{MacBouPouUpt24}
Rodrigo~Panosso Macedo, Patrick Bourg, Adam Pound, and Samuel~D. Upton.
\newblock Multidomain spectral method for self-force calculations.
\newblock {\em Physical Review D}, 110(8), October 2024.

\bibitem{CarHurVal}
Diego~A. Carranza, Adem~E. Hursit, and Juan A.~Valiente Kroon.
\newblock Conformal wave equations for the einstein-tracefree matter system.
\newblock {\em General Relativity and Gravitation}, 51(7), jul 2019.

\bibitem{Pae13}
T.-T. Paetz.
\newblock Conformally covariant systems of wave equations and their equivalence
  to {E}instein's field equations.
\newblock {\em Ann. Henri Poincar\'{e}}, 16:2059, 2013.

\bibitem{VanVal24}
Alex Va\~n\'o Vi\~nuales and Tiago Valente.
\newblock {Height-function-based 4D reference metrics for hyperboloidal
  evolution}.
\newblock {\em General Relativity and Gravitation}, 56(135), 11 2024.

\bibitem{Zen07a}
An{\i}l Zengino{\u{g}}lu.
\newblock Hyperboloidal foliations and scri-fixing.
\newblock {\em Classical and Quantum Gravity}, 25(14):145002, jun 2008.

\bibitem{Fra98a}
J.~Frauendiener.
\newblock Numerical treatment of the hyperboloidal initial value problem for
  the vacuum {E}instein equations. {II}. the evolution equations.
\newblock {\em Phys. Rev. D}, 58:064003, 1998.

\bibitem{Ste91}
J.~Stewart.
\newblock {\em Advanced general relativity}.
\newblock Cambridge University Press, 1991.

\bibitem{GasHil18}
Edgar Gasper\'in and David Hilditch.
\newblock {The Weak Null Condition in Free-evolution Schemes for Numerical
  Relativity: Dual Foliation GHG with Constraint Damping}.
\newblock {\em Class. Quant. Grav.}, 36(19):195016, 2019.

\bibitem{FraSte23}
J~Frauendiener and C~Stevens.
\newblock The non-linear perturbation of a black hole by gravitational waves.
  ii. quasinormal modes and the compactification problem.
\newblock {\em Classical and Quantum Gravity}, 40(12):125006, May 2023.

\bibitem{GarGasVal18}
A~García-Parrado Gómez-Lobo, E~Gasperín, and J~A Valiente~Kroon.
\newblock Conformal geodesics in spherically symmetric vacuum spacetimes with
  cosmological constant.
\newblock {\em Classical and Quantum Gravity}, 35(4):045002, January 2018.

\bibitem{Fri03a}
Helmut Friedrich.
\newblock Conformal geodesics on vacuum space-times.
\newblock {\em Communications in Mathematical Physics}, 235(3):513–543, April
  2003.

\bibitem{Hil16}
David Hilditch, Enno Harms, Marcus Bugner, Hannes R\"uter, and Bernd
  Br\"ugmann.
\newblock {The evolution of hyperboloidal data with the dual foliation
  formalism: Mathematical analysis and wave equation tests}.
\newblock {\em Class. Quant. Grav.}, 35(5):055003, 2018.

\bibitem{VanHusHil14}
Alex Va{\~n}{\'o}-Vi{\~n}uales, Sascha Husa, and David Hilditch.
\newblock {Spherical symmetry as a test case for unconstrained hyperboloidal
  evolution}.
\newblock {\em Class. Quant. Grav.}, 32(17):175010, 2015.

\bibitem{PanMac24}
Rodrigo Panosso~Macedo.
\newblock Hyperboloidal approach for static spherically symmetric spacetimes: a
  didactical introduction and applications in black-hole physics.
\newblock {\em Philosophical Transactions of the Royal Society A: Mathematical,
  Physical and Engineering Sciences}, 382(2267), January 2024.

\bibitem{RucEtiBau18}
Ian Ruchlin, Zachariah~B. Etienne, and Thomas~W. Baumgarte.
\newblock {SENR/NRPy+: Numerical Relativity in Singular Curvilinear Coordinate
  Systems}.
\newblock {\em Phys. Rev.}, D97(6):064036, 2018.

\bibitem{FerVicHil21}
Isabel Suárez~Fernández, Rodrigo Vicente, and David Hilditch.
\newblock Semilinear wave model for critical collapse.
\newblock {\em Physical Review D}, 103(4), February 2021.

\bibitem{BeyDouFraWha13}
Florian Beyer, Georgios Doulis, Jörg Frauendiener, and Ben Whale.
\newblock Linearized gravitational waves near space-like and null infinity,
  2013.

\bibitem{MacVal18}
Rodrigo~P Macedo and Juan~A Valiente~Kroon.
\newblock Spectral methods for the spin-2 equation near the cylinder at spatial
  infinity.
\newblock {\em Classical and Quantum Gravity}, 35(12):125007, May 2018.

\bibitem{Val03a}
J.~A. Valiente~Kroon.
\newblock Polyhomogeneous expansions close to null and spatial infinity.
\newblock In J.~Frauendiener and H.~Friedrich, editors, {\em The Conformal
  Structure of Spacetimes: Geometry, Numerics, Analysis}, Lecture Notes in
  Physics, page 135. Springer, 2002.


\end{thebibliography}
\end{document}